\documentstyle[epsf]{article}
\newcommand{\be}{\begin{equation}}
\newcommand{\ee}{\end{equation}}
\newcommand{\bea}{\begin{eqnarray}}
\newcommand{\eea}{\end{eqnarray}}
\newcommand{\beas}{\begin{eqnarray*}}
\newcommand{\eeas}{\end{eqnarray*}}
\newcommand{\bookfig}[5]{
\begin{figure}\centering\fbox{\epsfysize=#5cm \epsfbox{#1}}
\caption[#2]{#4}\label{#3}
\end{figure}
}

\begin{document}
\title{Chen's Iterated Integral represents the Operator
Product Expansion}
\author{D.~KREIMER\thanks{Heisenberg Fellow}\\
\small Dept.~of Physics, Mainz Univ.\\
\small  D-55099 Mainz, Germany\\
\footnotesize dirk.kreimer@uni-mainz.de}
\date{\small {\bf MZ-TH/98-50}, Jan.21 1999}
\maketitle


\begin{abstract}
The recently discovered formalism underlying renormalization
theory, the Hopf algebra of rooted trees, allows to generalize
Chen's lemma. In its generalized
form it describes the change of a scale in Green functions, and hence relates
to the operator product expansion. Hand in hand with this
generalization goes the generalization of the ordinary factorial $n!$
to the tree factorial $t^!$.
Various identities on tree-factorials are derived which clarify
the relation between Connes-Moscovici weights and Quantum Field Theory.
\end{abstract}
\section{Introduction}
In this paper we want to explore a close resemblance between a
mathematical structure, iterated integrals \cite{ktC}, and a
structure from physics, renormalization. Renormalization theory
has been recently identified to be rooted in a Hopf algebra
structure, which encapsulates its combinatorial properties
\cite{hopf}. Further, renormalization establishes a calculus which
generalizes the algebraic approach to the diffeomorphism group,
featured by Connes and Moscovici \cite{CK,CM}. Structures due to
this generalization are relevant for the practitioner of
perturbative Quantum Field Theory  (pQFT) \cite{DiDa}.


Here, we will study all of these aspects in some detail, featuring
in particular the role of renormalization schemes, the
renormalization group and the operator product expansion (OPE).


To proceed, we push forward two parallel developments.
One is the use of toy models at a more and more sophisticated level, which provide
clarifying examples to exhibit underlying ideas on which the reader can
check formal developments, which will establish a set-up which allows to
recover the standard notions established by physicistes, the before mentioned
existence of renormalization schemes, renormalization groups and
operator product expansions.


In this manner, we will show that the Hopf algebra of rooted trees not only
describes the combinatorics of renormalization,
but also analytical structure: the behaviour under variations of scales.


Throughout this paper we assume that the results and notions of
\cite{hopf,CK,DiDa,overl} are familiar.
We will nevertheless summarize some basic notions and conventions.
\subsection{Notation}
The Hopf algebra of rooted trees (which are possibly decorated
multiplicative generators) is denoted by ${\cal H}$, with
coassociative coproduct $\Delta$, antipode $S$ and counit
$\bar{e}$. The multiplication in the algebra is denoted by $m$. It
is commutative, and hence $S^2=1$. The unit of the algebra is
denoted by $e$. The counit is denoted by $\bar{e}$, with
$\bar{e}(e)=1$ and $0$ otherwise.
The underlying number-field is
assumed to be ${\bf Q}$.


The fertility of a vertex of a rooted tree
is the number of outgoing edges. The root
is always drawn as the uppermost vertex, and all edges are oriented
away from the root.


Rooted trees $t$ are graded by
their number of vertices $\#(t)$.
For a product of rooted trees
$\prod_i t_i$ we define
$\#(\prod_i t_i)=\sum_i \#(t_i)$. Obviously, $\#(e)=0$.


We abbreviate the coproduct using Sweedler's notation:
$\Delta(t)=\sum t_{(1)}\otimes t_{(2)}$ $\forall t\in {\cal H}$.
\subsection{Summary of sections}
This paper is organized as follows.
In section two we introduce the iterated integral.
In particular, we focus on iterated integrals which have
a divergence at the upper boundary, which is a choice
motivated by the renormalization problem in QFT,
formulated in momentum space.
We show how we get well-defined iterated integrals
as renormalized Green functions, and show that a variation
of scales amounts to an application of Chen's Lemma.


In section three we focus on the multiplicativity of renormalization.
Crucial is the formulation of multiplicativity constraints,
which are sufficient to derive the multiplicativity
of counterterms. We present a boundary operator $d_R$
for any renormalization scheme $R$ and show that all
renormalization schemes can be treated on the same footing on the
expense of introducing tree-indexed parameters.


Section four applies those results to a restricted class of Feynman
diagrams, those which represent trees with the same decoration
at each vertex. Such classes were considered already in \cite{DiDa}.
Here, we use them to exemplify the results of section three.


Section five proves some identities for rooted trees which were
conjectured in \cite{DiDa} and which are useful in understanding the
relation to noncommutative geometry.


Section six gives the principal reason why operator product expansions
are related to Chen's Lemma.


Conclusions finish the paper. It is the main objective of this
paper to introduce and exhibit some essential properties of the
Hopf algebra approach underlying renormalization, using iterated
integrals as a convenient toy all along on which the reader can
test the relevant  notions. The translation to proper Green
functions is a notational exercise which can be conveniently
spelled out whenever needed, as, for example, in \cite{DKnew}.


While this paper introduces essential conceptual properties, details
will be presented in future work.
\section{Iterated Integrals and Renormalization}
The crucial feature of renormalization is the fact that it is governed
by its underlying Hopf algebra structure of rooted trees.
Bare Green functions as they appear in a perturbative
approach to  QFT based on polynomial interactions provide a representation
of this Hopf algebra. Apart from a systematic access to the renormalization
problem of such QFTs the Hopf algebra also allows to study other representations
and hence to define models for the renormalization problem which deliver handy tools
to study more advanced topics, for example
the change of scales and renormalization schemes.
In this section, we will largely consider iterated integrals
for that purpose.
\subsection{The iterated integral}
We will start our considerations by reminding ourselves of some basic
properties of iterated integrals \cite{ktC,ShnSt}. We specialize
to the case of a single function $f(x)$ with associated
one-form $f(x)dx$ on the real line.


Then, iterated integrals built with the help of
$f$ are parametrized by an integer $n$, and two real numbers $a,T$ say.
They are defined by
\bea
F^{[0]}_{a,T} & = & 1,\,\forall a,T\in {\bf R},\label{triv}\\
F^{[n]}_{a,T} & = & \int_a^T f(x)F^{[n-1]}_{a,x}dx,\,\forall n>0.
\eea
Hence we can write them as an integral over the simplex
\be
F_{a,T}^{[n]}=\int_{a\leq x_1<\ldots<x_{n}\leq T}
f(x_1)\ldots f(x_n) dx_1\ldots dx_n.
\ee
We can easily generalize this to the case of different functions $f_i$,
and, defining a string of integers $I=(i_1,\ldots, i_n)$,
we can define
\be
F_{a,T}^{I}=
\int_{a\leq x_1<\ldots<x_{n-1}\leq x_n} f_{i_1}(x_1)
\ldots f_{i_n}(x_n) dx_1\ldots dx_n,
\ee
where $i_k$ are integers, taken from some
index set ${\cal I}$, labelling the available forms
$f_{i_k}(x_k)dx_k$.


It is a well-known fact \cite{ShnSt} that such integrals fulfil a convolution
\be
F_{a,T}^{I}=F_{a,s}^I+F_{s,T}^I+
\sum_{I=(I^\prime I^{\prime\prime})}F_{a,s}^{I^\prime}
F_{s,T}^{I^{\prime\prime}},\label{chen}
\ee
where the sum is over the $n-1$ partitions of the string
$I$ into two non-empty substrings $I^\prime,I^{\prime\prime}$.
We shall dubb (\ref{chen}) Chen's Lemma, following \cite{ShnSt}.
\subsection{Renormalization of iterated integrals}
Let us motivate our interest in iterated integrals and Chen's Lemma.
Consider again the trivial case of only one $f$ and let us assume
it behaves for large $x\gg b$ as $f(x)\equiv f(\epsilon;x)\sim x^{-1-\epsilon}$,
for $0<\epsilon\ll 1$.
Let us then define
\be
G^{[t_2]}_{b,\infty}
=
\left[ \int_b^\infty \left( \int_x^\infty f(y) dy\right)
f(x)dx\right]. \ee Then, in the limit $\epsilon\to 0$, this
expression is ill-defined. It has the structure of a nested
$y$-integration, furnishing a subdivergence (in $()$-brackets) in
the jargon of renormalization theory, which is nested inside the
final $x$-integration, which diverges as well and thus provides an
overall divergence (in $[]$-brackets).


To such a combination of ill-defined
integrations  we  associate a rooted tree $t=t_2(f,f)$, as in Fig.(\ref{f1}),
following the guidance of \cite{hopf,CK}.
Generalizing to arbitrary $n$, we define, $\forall b\in {\bf R}_+$,
functions
\bea
G^{[e]}_{b,\infty} & = & 1,\\
G^{[t_n]}_{b,\infty}
 & = & -
\left[ \int_b^\infty \left( G^{[t_{n-1}]}_{x,\infty} \right)
f(x)dx\right],\;\forall n\geq 1. \eea To them, we assign the
rooted tree $t_n:=B_+^n(e)$ of $n$ vertices without
side-branching, as in Fig.(\ref{f1}), and understand that the
empty tree, the unit $e$ of the Hopf algebra ${\cal H}$, is
associated to $G^{[e]}_{b,\infty}=1$. As a decorated rooted tree
$t_n$ carries the same decoration $f$ at each vertex.
\bookfig{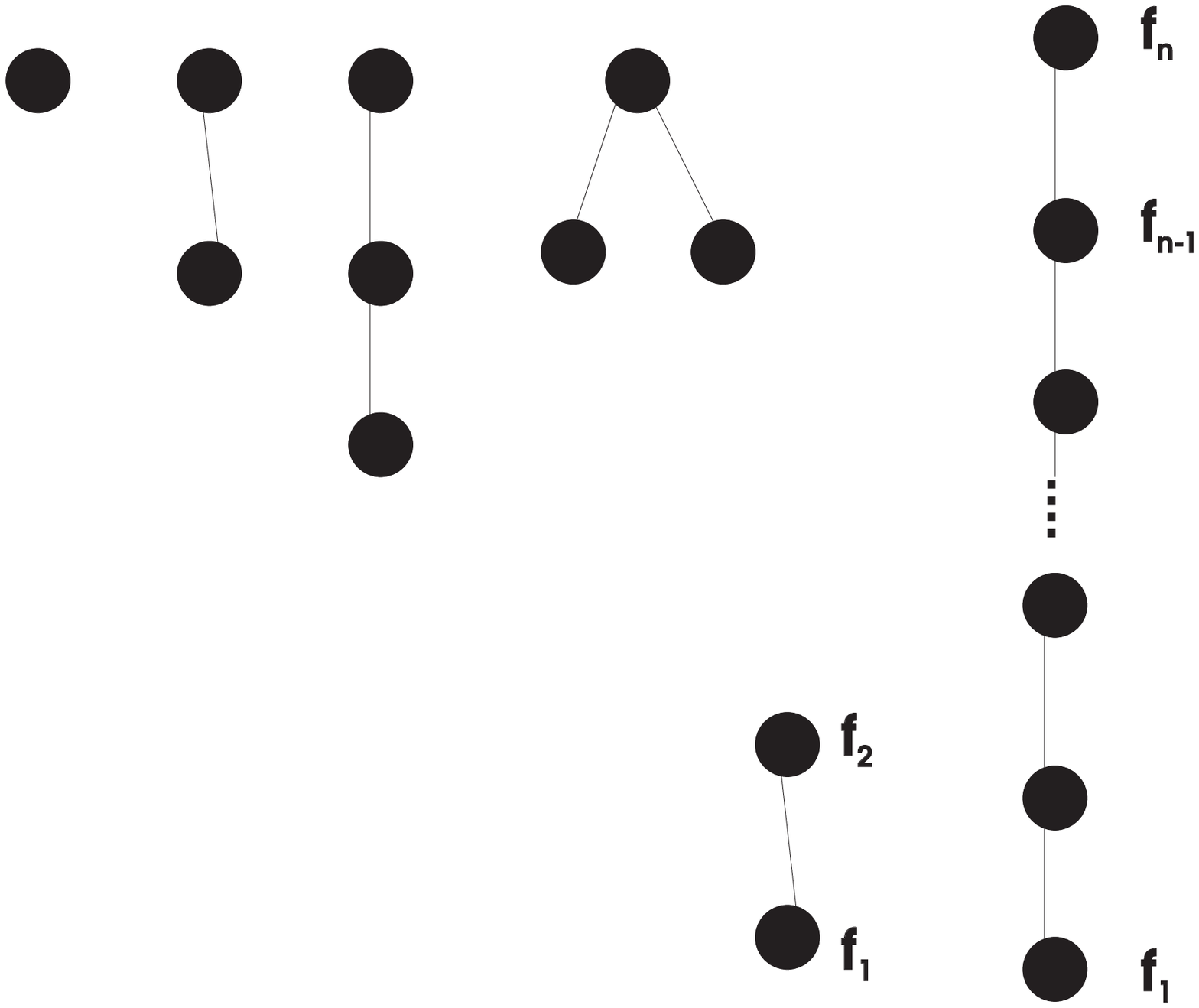}{rt}{f1}{Rooted trees describe nested
integrations. We give the trees $t_1,t_2,t_{3_1},t_{3_2}$ from
left to right and a decorated tree without sidebranchings,
$B_+^n(e)$, with decorations $f_n$ to $f_1$. Also, we explicitly
give the tree $t_2(f_1,f_2)$ for the case $n=2$.}{5}


It is straightforward to see that the trees $t_n$ form a closed
sub-Hopf algebra ${\cal H}_{Chen}$ of ${\cal H}$, which is a Hopf
algebra based on rooted trees without sidebranchings: \bea
\Delta[t_n] & = & t_n\otimes e+e\otimes
t_n+\sum_{i=1}^{n-1}t_i\otimes t_{n-i}\\ S[t_n] & = &
-t_n-\sum_{i=1}^{n-1} S[t_i]t_{n-i}. \eea This remains true for
decorated rooted trees and in the same spirit, we can assign a
decorated rooted tree $t_I$ to any function
\be
G^{[t_I]}_{b,\infty}= -\int_b^\infty G^{[t_I^\prime]}_{x,\infty}f_{i_n}(x)dx,
\ee
where $t_I^\prime$ is the rooted tree
$B_-(t_I)$, providing an index string  which has  the $n$-th entry $i_n$
of $I$ deleted. The Hopf algebra of decorated rooted trees
without sidebranchings is still denoted by ${\cal H}_{Chen}$.


Note that the functions $G^{[t_I]}_{b,\infty}$ can be regarded as
iterated integrals in their own right:
\be
G^{[t_I]}_{b,\infty}=\lim_{\lambda\to\infty}F^I_{\lambda,b}.
\ee


All the functions $f_i$, $i\in {\bf N}$, are assumed to behave as
$\lim_{x\to\infty}f_i(x)=c_i x^{j_i-\epsilon}$, for some constant
$c_i$ and some integer $j_i$ with $j_i\geq -1$. Hence, in the
limit $\epsilon\to 0$, these iterated integrals are ill-defined,
due to a divergence at the upper boundary. We will have to
renormalize them. The Hopf algebra will allow us to find a way to
make sense out of the expressions $G^{[t_I]}_{b,\infty}$ at
$\epsilon=0$. We will discuss further aspects of the behaviour at
infinity in some detail later. All renormalization properties
discussed below extend in an obvious way to the renormalization at
endpoints different from infinity, if it so happens that the
functions $f_i(x)$ have singularities at such endpoints. The
renormalization procedure is a very natural operation, as we will
see, and one can define applications largely extending the task of
eliminating UV divergences. A recent review on its relations to
many branches of science can be found in \cite{CK2}. An obvious
application is to the configuration space of $n$ distinct points,
which  can be tested out by differential forms which diverge at
(sub-)diagonals. Such an application will be described in
\cite{CKnew}.


We can multiply the functions and add the functions
$G^{[t_I]}_{b,\infty}$ freely. Hence,
if $\phi:$ ${\bf R_+}\times{\cal H}_{Chen}\to
V$ is the map which assigns to any decorated rooted tree
$t_I\in{\cal H}_{Chen}$ and positive real number $b$
the function $G^{[t_I]}_{b,\infty}$, then this gives us
a representation,
parametrized by $b$,
\be
\phi(b;t_I t_J)=\phi(b;t_I)\phi(b;t_J). \ee We also set
$\phi(b;0)=0$ and $\phi(b;e)=1$, $\forall b$, in accordance with
(\ref{triv}). For a chosen $b$, we further write $\phi_b: {\cal
H}_{Chen}\to V$, $t\to \phi_b(t):=\phi(b;t)$. The target space $V$
can be considered as the ring ${\bf R}[\epsilon^{-1},[\epsilon]]$
of Laurent series with poles of finite order. The parameter $b$ is
from now on denoted as the scale of the representation. The
functions $\phi(b,t)=G^{[t]}_{b,\infty}$ can be considered as role
models for bare Green functions. They depend on an external scale
$b$. We will utilize this dependence to define the renormalization
procedure.


Let $R_a$ be the map which sends $\phi_b\to \phi_a$. We thus evaluate
at a different scale.
Essentialy, we claim, it is this change of external scale(s)
which allows us to renormalize
in a non-trivial manner. To the  bare Green function $\phi_b:{\cal H}_{Chen}\to V$,
which defines a representation of ${\cal H}_{Chen}$, we associate
another function $S_{R_a}(\phi_b):{\cal H}_{Chen}\to V$ by
\be
S_{R_a}(\phi_b)(t):=-R_a[\phi_b(t)+m[(S_{R_a}\otimes
id)(\phi_b\otimes \phi_b)P_2\Delta(t)]], \ee defined for any
monomial $t$ of decorated rooted trees,  $t\not=e$. For $t=e$ we
set $S_{R_a}(\phi_b)(e)=1$. In the above, $P_2$ denotes the
projector $(id-e\bar{e})\otimes(id-e\bar{e})$ which annihilates
any appearance of the unit $e$.

The resulting map $S_{R_a}(\phi_b)$ is independent of $b$ by the
definition of $R_a$ which eliminates any dependence on the scale
$b$. Two examples, $t=t_1(f_i)$ and $t=t_2(f_i,f_j)$ might be
useful: \bea S_{R_a}(\phi_b)(t_1(f_i)) & = &
-\phi_a(t_1(f_i))=\int_a^\infty f_i(x)dx,\\
S_{R_a}(\phi_b)(t_2(f_i,f_j)) & = &
-\phi_a(t_2(f_i,f_j))+\phi_a(t_1(f_i))\phi_a(t_1(f_j)) \nonumber\\
 & = &
-\int_a^\infty \int_x^\infty f_i(y)dy f_j(x)dx+\int_a^\infty f_i(x)dx\int_a^\infty
f_j(y)dy\nonumber\\
 &= &
\int_a^\infty \int_a^x f_i(y)dy f_j(x)dx.
\eea
Now, we define a function $\Gamma:{\bf R_+}\times{\bf R_+}\times {\cal H}_{Chen}
\to V$ by
\bea
\Gamma_{a,b}(t) & = &
\sum S_{R_a}(\phi_b)(t_{(1)})\;\phi_b(t_{(2)})\label{gam1}\\
 & = & m[(S_{R_a}\otimes id)(\phi_b\otimes \phi_b)\Delta(t)]\label{gam2}\\
 & = & m[(\phi_{R_a}\otimes \phi_b)(S\otimes id)\Delta(t)],\label{gam3}
\eea where we define $\phi_{R_a}:{\cal H}_{Chen}\to V$ by
$\phi_{R_a}=S_{R_a}(\phi_b\circ S)$. Note that this renormalized
function has naturally the structure of a ratio, comparing to
scalar functions of rooted trees with the help of the antipode.
This has far reaching consequences \cite{Brou,CK2}.


$\phi_{R_a}$ is still independent of $b$ by the definition of
$R_a$. Note further that $\Gamma_{a,b}(t)$ exists in the limit
$\epsilon\to 0$ when we integrate to infinity and can be regarded
as the renormalized iterated integral associated to the bare
iterated integral $G_{b,\infty}^t$. The equality between
(\ref{gam2}) and (\ref{gam3}) follows because of
$S_{R_a}\circ\phi_b=S_{R_a}\circ\phi_b\circ S^2=\phi_{R_a}\circ
S$,using the definition of $\phi_{R_a}$ and $S^2=id$.\\ {\bf
Prop.1}: $\Gamma_{a,b}(t)=F_{a,b}^I,$\\ where $t$ is the decorated
rooted tree with $n$ vertices corresponding to the string $I$.\\
Proof: Straightforward (for example, use (\ref{chen}) and that
$G^t_{b,\infty}$ is itself an iterated integral from $b$ to
$\infty$).~$\Box$\\ Example: \bea \Gamma_{a,b}(t(f_i)) & = &
\left[ -\int_b^\infty+\int_a^\infty \right]f_i(x)dx= \int_a^b
f_i(x)dx,\\ \Gamma_{a,b}(t(f_i,f_j)) & = & \left[
\int_b^\infty\int_x^\infty-\int_b^\infty\int_a^\infty
-\int_a^\infty\int_x^\infty+\int_a^\infty\int_a^\infty \right]
f_i(y)dyf_j(x)dx\nonumber\\
 & = & \int_a^b\int_a^x f_i(y)dy f_j(x)dx.
\eea
We now read this as an instructive example for renormalization.
The role of a bare Green function, demanding renormalization,
is played by $G_{b,\infty}^t\equiv \phi_b(t)$. It provides
$n-1$ subdivergences, as all integrations diverge at the upper boundary.


Then, $S_{R_a}[\phi(t_I)]$ delivers a counterterm such that
$\Gamma_{a,b}(t_I)$ is a quantity which is renormalized: it contains
only well-defined integrations and the limit $\epsilon\to 0$
can be taken at the level of integrands.


Note that if the bare Green function would be independent
of the external scale furnished by the parameter $b$, then
$\phi_a(t)=\phi_b(t)$ and as a consequence,
$\Gamma_{a,b}(t)\equiv 0$.\footnote{This makes dimensional regularization
a succesful regularization scheme in practice: it annihilates
scale independent terms from the beginning, and is hence extremely economic.}


This can be utilized to show that $\Gamma_{a,b}(t)$ is determined
by the coefficient of logarithmic divergence at infinity. It is
thus natural to look for a representation of ${\cal H}$ in terms
of residues in the sense of \cite{ACbook,CK2}.




For the moment, we note that the presence of a second scale $a$
is unavoidable if we want to go from bare functions to renormalized
ones.  $\Gamma_{a,b}(t)$ is essentially the ratio of two
representations, one
parametrized by $b$, the other by $a$. We call $a$ the renormalization point.
For all possible values of the external scale $b$ there is one place,
the diagonal $b=a$, at which $\Gamma_{b,b}(t)=0$. Further,
\be
R_a(\Gamma_{a,b}(t))=R_a[(S_{R_a}\star id)\phi_b](t)=\bar{e}(t),
\ee showing that $S_{R_a}$ is the inverse of the identity in the
range $V_{R_a}$ of $R_a$ in $V$. This inverse is taken  with
respect to the induced convolution in ${\cal
H}_{Chen}^\star\otimes V$
\be
[\psi\star\phi](t)=\sum\psi(t_{(1)})\phi(t_{(2)}),
\ee
valid for all maps $\psi,\phi:{\cal H}_{Chen}\to V$.



Let us summarize: We start with a bunch of ill-defined integrations,
labelled by a decorated rooted tree $t$ from which we can determine
the bare integral demanding renormalization.


We then construct the analytic expressions determined by the
counterterm map $S_{R_a}:{\cal H}_{Chen}^\star\otimes V\to
{\cal H}_{Chen}^\star\otimes V$. This gives rise to a renormalized
iterated integral $\Gamma_{a,b}(t)$
which only involves well-defined integrations. It assigns a well-defined
analytic expression to any decorated rooted tree.
This expression necessary vanishes along the diagonal $a=b$.
In contrast to this, $\phi_b$ associated the ill-defined bare
integral $\phi_b(t)$ to any rooted tree $t$. This transition from $\phi_b(t)$
to $\Gamma_{a,b}(t)$ is what renormalization
typically achieves.


A final remark in this section concerns the solution to the
Knisz\-hnik-Zamo\-lodchi\-khov (K-Z) equation in two variables, based on forms
$\frac{dz}{z},\frac{dz}{1-z}$, say.
See  \cite{ShnSt} for a review.


Consider the
K-Z equation
\be
\frac{dF}{dz}= \left( \frac{a}{z}-\frac{b}{1-z}\right)F.
\ee
Here, $a$ and $b$ are two {\em noncommuting} variables which
actually provide a free Lie algebra on two elements.
Arbitrary words out of the two-letter alphabet
$\{a,b\}$
are considered and no relation amongst such words exist.
Let $W$ be the set of all words.


The length of such a word $w$  is $l(w)$
and the $i$'th letter of $w$ is  $w(i)$.
Let us consider the following expression
\bea
G(u,v) & = &
\sum_{w \in W} \int_{\Delta(u,v)} \prod_{i: w(i)=a}
\frac{dz_i}{z_i}\prod_{i: w(i)=b}\frac{dz_i}{(1-z_i)}\\
 & & \Delta(u,v)=u>z_{l(w)}>\ldots>z_1>v.\label{guv}
\eea
$G(u,v)$ is known to be a solution to the K-Z equation in the interval $]0,1[$.


$G(u,v)$ contains multiple zeta values \cite{prog} (MZV's) for $(u,v)=(1,0)$,
whenever the limits $u\to 1$, $v\to 0$ are defined.
But whenever a word starts with $b$ or ends with $a$ these limits do not exist.


Hence this solution is a series $\sum_w w F_{v,u}^w$ over all words
built out of two noncommuting variables $a,b$ multiplying
iterated integrals $F_{v,u}^w$ in the interval
$]0,1[$ which possibly diverge at both endpoints of the interval
$]0,1[$.
Let $W^\prime$ be the words which neither start with $b$ nor end with $a$.


Renormalization can be applied to $G(u,v)$ such that it
extends to  either boundary,
and applying it successively using the renormalization schemes
$R_0$ (which removes all words ending with $a$)
and $R_1$ (which removes all words beginning with $b$)
at the lower and upper boundaries
leaves us with the K-Z associator
\be
\phi_{KZ}=\sum_{w\in W^\prime}
\int_{\Delta(u,v)} \prod_{i: w(i)=a}
\frac{dz_i}{z_i}\prod_{i: w(i)=b}\frac{dz_i}{(1-z_i)}
\ee
as the renormalized 'Green function'.
An iterated integrals accompanying the
word $b^{m_1}a^{n_1}b^{m_2}\ldots a^{n_{k-1}}b^{m_k}a^{n_k}$
must be regarded as a representative of
 the rooted tree $B_+^{m_1}(e)$ for the renormalization
at the upper boundary $1$ (where $b$ is the variable assigned to $dz/(1-z)$).
For the renormalization
at the lower boundary it represents $B_+^{n_k}(e)$
(where $a$ is the variable assigned to $dz/z$).
Then, the renormalized iterated integral assigned to $G(u,v)$
extends to $[0,1]$, where
it is the above K-Z associator.
\subsection{Change of Scales}
It is most interesting to consider the behaviour if we change the
renormalization point $a\to a^\prime$, which will lead us to the
group law underlying the evolution of functions of rooted trees
quite generally, which is the group law of the Butcher group
(comp.~\cite{Brou,CK2} and references there). One gets
\be
\Gamma_{a,b}(t_I)=\sum_{I=(I^\prime,I^{\prime\prime})}\Gamma_{a,a^\prime}({t_{I^\prime}})
\Gamma_{a^\prime,b}(t_{I^{\prime\prime}}).\label{gchen0} \ee
Proof: this is just (\ref{chen}) for iterated integrals
\cite{ShnSt}. Nevertheless, let us derive it from the Hopf algebra
structure of ${\cal H}$. At this stage we should actually use
${\cal H}_{Chen}$. But as we will see that nothing in the
following derivation depends on the peculiarities of this sub-Hopf
algebra of ${\cal H}$, we directly use the latter instead. Hence
define the following operator
\be
U:{\cal H}\otimes {\cal H}\otimes {\cal H}
\otimes {\cal H}\to V
\ee
\be
U=
m[m\otimes m](\phi_{R_a}\otimes\phi_{a^\prime}
\otimes\phi_{R_{a^\prime}}\otimes \phi_b).
\ee
Composition with $M:$ ${\cal H}\to{\cal H}\otimes{\cal H}\otimes{\cal H}\otimes{\cal H}$,
\bea
M: & = & (S\otimes id\otimes S\otimes id)(\Delta\otimes\Delta)\Delta\nonumber\\
 & = & (S\otimes id\otimes S\otimes id)(\Delta\otimes id\otimes id)(id\otimes\Delta)\Delta
\nonumber\\
 & = & (S\otimes id\otimes S\otimes id)(id \otimes \Delta\otimes id)(id\otimes\Delta)\Delta
\eea
gives
\[
U[M(t)]=\phi_{R_a}[S(t_{(1)})]\phi_{a^\prime}[t_{(2)}]\phi_{R_{a^\prime}}[S(t_{(3)})]\phi_b[t_{(4)}],
\]
where coassociativity of ${\cal H}_R$ allows to use Sweedler's
notation throughout. The above is \bea
\phi_{R_a}[S(t_{(1)})]m[(\phi_{a^\prime}\otimes\phi_{R_{a^\prime}})
(id \otimes S)\Delta(t_{(2)})]\phi_b(t_{(3)}). \eea As
\[
m[(\phi_{a^\prime}\otimes\phi_{R_{a^\prime}})(id\otimes S)\Delta(t)]=
\phi_{a^\prime}(m[(id\otimes S)\Delta(t)])=
\phi_{a^\prime}(\bar{e}(t))=\phi_{a^\prime}(0)=0,
\]
$\forall t\not=1$,
this has contributions only for $t^{(2)}=e$, in which case we obtain
\be
m[(\phi_{R_a}\otimes\phi_b)(S\otimes id)\Delta(t)], \ee as
desired. We used $S_{R_{a^\prime}}(\phi_b)=\phi_{a^\prime}\circ
S$, which we prove later, see (\ref{cha}).~$\Box$\\


Nothing in this derivation prevents us to generalize to arbitrary rooted trees,
extending from ${\cal H}_{Chen}$ to ${\cal H}$.
We thus define, for a decorated rooted tree $t\in {\cal H}$ with $n$ vertices,
\be
G_{b,\infty}^{t}=-\int_b^\infty f_{i_n}(x) \prod_j G_{x,\infty}^{[t^\prime]_j}dx,
\ee
where the product
is over the decorated branches of the decorated
tree $t$, $B_-(t)=\prod_j t^\prime_j$ and $f_{i_n}(x)$ is the label
attached to the root. We still write $\phi_b(t)$ for $G_{b,\infty}^t$,
but stress that $\phi_b(t):{\cal H}\to V$
now gives parametrized representation for the full Hopf algebra ${\cal H}$
of decorated rooted trees.


We also define, $\forall t\not= e$, the functions
$S_{R_a}(\phi_b)(t)$ and $\Gamma_{a,b}(t)$ without any change:
\be
S_{R_a}(\phi_b)(t)=-R_a[\phi_b(t)+m[(S_{R_a}\otimes
id)(\phi_b\otimes\phi_b)P_2\Delta(t)]] \ee and \bea
\Gamma_{a,b}(t) & = & \sum
S_{R_a}(\phi_b(t_{(1)}))\phi_b(t_{(2)})\\
  & = & m[(S_{R_a}\otimes id)(\phi_b\otimes \phi_b)\Delta(t)]\\
 & = & m[(\phi_{R_a}\otimes \phi_b)(S\otimes id)\Delta(t)].
\eea


Then, in a straightforward generalization
one concludes from the above derivation\\
{\bf Lemma 1:}
\be
\Gamma_{a,b}(t)=[S_{R_a}(\phi_b)\star\phi_b](t)=[\phi_{R_a}\circ
S\star\phi_b](t)=
\sum\Gamma_{a,a^\prime}(t_{(1)})\Gamma_{a^\prime,b}(
t_{(2)}).\label{gchen} \ee This lemma holds for any scalar
function of rooted trees (with generalizations to matrix functions
worked out in \cite{DKnew}) and hence applies to full QFT
\cite{DiDa} as well.

Example: Let $t$ be the decorated rooted tree of Fig.(\ref{f2}).
Then, $G_{b,\infty}^{t}$ is given as
\be
G_{b,\infty}^{t}= \int_b^\infty f_1(x_1)\int_{x_1}^\infty f_2(x_2)
\int_{x_2}^\infty f_1(x_3)dx_3\int_{x_2}^\infty f_3(x_4)dx_4 dx_2
dx_1. \ee Accordingly, $\Gamma_{a,b}(t)$ becomes \bea
\Gamma_{a,b}(t) & = & \underbrace{ \int_a^b f_1(x_1) \underbrace{
\int_{a}^{x_1} f_2(x_2) \underbrace{ \int_{a}^{x_2} f_1(x_3)dx_3}
\underbrace{ \int^{x_2}_a f_3(x_4)dx_4} dx_2} dx_1}, \eea as the
reader should check. The underbracings indicate the tree structure
of the nested and disjoint subintegrations, which is also
exemplified in Fig.(\ref{f2}). \bookfig{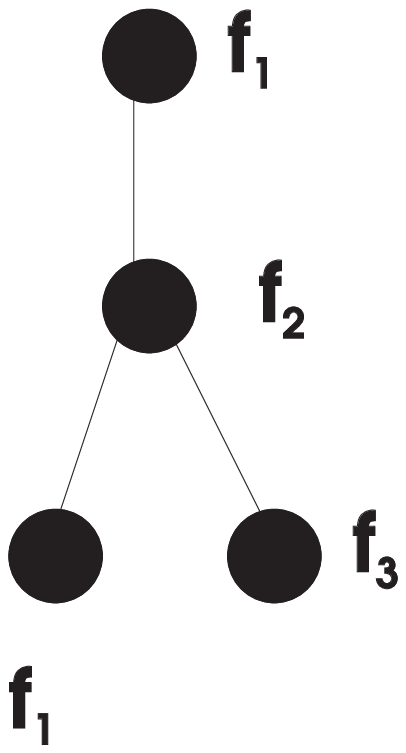}{rtw}{f2}{An
example.}{3}


A few remarks. We obtain a natural generalization of the iterated
integral. Actually, due to the fact that iterated integrals obey
the shuffle product, it is not yet a generalization, as any bare
integral representing a tree  with side branchings is a linear
combination of integrals representing trees in ${\cal H}_{Chen}$.
Hence, at this stage, the generalization is merely  a convenient
notation. But there are more generalizations lying ahead,
considering representations of ${\cal H}$ which extend the notions
of iterated integrals truely, still obeying the convolution which
implies Chen's lemma for ordinary iterated integrals. In
particular, bare Green functions as typically derived from Feynman
rules in the perturbative approach to a local QFT represent rooted
trees in a manner such that trees with sidebranchings cannot be
reexpressed in terms of trees without sidebranchings, as a shuffle
product is absent in such circumstances.\footnote{Though, as
reported elsewhere \cite{DKnew}, some remainders of it are still
visible in QFT.} Nevertheless, the derivation of the convolution
of renormalized functions is a mere application of the
coassociativity of ${\cal H}$ itself, and hence applies whenever
appropriate representations of the Hopf algebra are available.


It will be an interesting exercise in the future to understand the
monodromy of Green functions from this approach in the same manner
as one can understand the
monodromy of the polylogarithm from the study of
the renormalized solution of the K-Z equation.
Green functions in QFT are a more general class of
functions than polylogs.
Nevertheless, at lower loop orders, they
 are intimately related which
might well be understood
one day as testimony to the fact that Green functions
in pQFT  realize in a wider set-up algebraic structures which polylogs
strictly obey.



Another generalization lies in the possibility to consider tree-indexed
parameters in the integral. This will turn out a convenient means
to parametrize the freedom which we have in the renormalization approach. To this
idea we come back soon.


It is an interesting question  which information
about a manifold $M$ such  generalizations
can provide, using as functions $f_i$
the pull-back of appropriate forms $\omega_i$ via paths on that manifold,
or, vice versa, how one can construct manifolds providing, eventually,
iterated integrals which evaluate to the same renormalized
Green functions as a QFT.
This amounts to setting up and solving appropriate systems of equations,
making use of the recursive properties of the Hopf algebra.
Apart from a few further remarks along these lines in the next section the reader will find
examples in \cite{DKnew}.


Before we consider further generalizations by tree-indexed scales,
we come to  some interesting
structures which can be readily observed at this level.
\section{Multiplicativity of renormalization and consequences}
So far, we observed how a change in the renormalization
point is expressed by the generalized form of Chen's Lemma.
This gives a very nice handle on the renormalization group (see below),
and relates it to quite general algebraic considerations.
While in this paper we will only outline the basic concepts,
concrete applications will be worked out in future work. Also, in \cite{DiDa},
the reader already finds applications which prove the usefulness of the
reduction of renormalization concepts to the Hopf algebraic set-up.


\subsection{Multiplicativity}
Remarkable features appear when one engulfes in a detailed study of the properties
under a change of renormalization schemes.
To this end, let us come back to the map $\phi_b:{\cal H}\to V$.
Clearly, $\forall t\not= e$,
\be
0=\phi_b(0)=\phi_b(\bar{e}(t))=\phi_b(m[(S\otimes id)\Delta(t)])=
m[(\phi_b\otimes\phi_b)(S\otimes id)\Delta(t)].
\ee
Compare the expression on the rhs with the expression for $\Gamma_{a,b}(t)$
\be
\Gamma_{a,b}(t)=m[(\phi_{R_a}\otimes\phi_b)(S\otimes id)\Delta(t)].
\ee
Hence, this expression is non-vanishing only
because of $S_{R_a}\circ \phi_b\circ S\equiv\phi_{R_a}\not=\phi_b$,
hence, essentially only if $a\not= b$.\footnote{One has
$\phi_{R_b}=\phi_b$, as one immediately checks.}
There is a map  $\Delta_b:{\cal H}\to V\otimes V$ induced by $\phi_b$,
\be
\Delta_b=(\phi_b\otimes\phi_b)\circ\Delta,
\ee
and an induced map  $S_b:{\cal H}\to V$, $S_b=\phi_b\circ S$.
It is more interesting to consider
the map ${\cal R}:{\cal H}^\star\otimes V\to
{\cal H}^\star\otimes V$ given by ${\cal R}(\phi)=S_R[\phi\circ S]$ defined
for any $\phi:{\cal H}\to V$ (hence, for any $\phi\in {\cal H}^\star\otimes V$).
We will now show that ${\cal R}({\cal R}(\phi))={\cal R}(\phi)$,
which is the natural extension of $R^2=R$.
This will allow us to define beautiful cohomological properties
for renormalization. We start with the consideration of
renormalization schemes which merely vary external parameters.
In the following, the reader should have in mind that $a,b$ are to be considered
as representatives of appropriate sets of external parameters (masses, external momenta
in Green functions) which parametrize analytic expressions representing elements
of ${\cal H}$.


It suffices to show
\be
S_{R_a}(\phi_b)(t)=\phi_a(S(t)),\label{cha}
\ee
which one readily proves
by induction on the number of vertices of $t$:
\bea
S_{R_a}(\phi_b)(t) & = &
-R_a[\phi_b(t)+m[(S_{R_a}\otimes id)(\phi_b\otimes\phi_b)
P_2\Delta(t)]]\\
 & = & -\phi_a(t)-R_a[m[(\phi_a\otimes\phi_b)(S\otimes id)P_2\Delta(t)]]\\
 & = & \phi_a\left(-t-m[(S\otimes id)P_2\Delta(t)]\right)\\
 & = & \phi_a(S(t)).
\eea
In the second line, we used that
\be
R_a[\phi_a(t)\phi_b(t^\prime)]=
R_a[\phi_a(t)]R_a[\phi_b(t^\prime)],\;\forall t,t^\prime\in{\cal H},
\ee
an equation which is fulfilled by $R_a$, but not by general renormalization maps.


If we regard a renormalization map $R:V\to V$ as simply a map from $V$ (considered
as a vector space) to $V$, then, in general,
$R[xy]\not= R[x]R[y]$, $\forall x,y\in V$.


A good example is a minimal subtraction scheme, which we will discuss in some
detail below. We can define its renormalization map
$R_{MS}$  by a projection to the pole part: if
\be
1\not=v =\sum_{i:=0}^\infty c_{-k+i}\epsilon^{-k+i}\in{\bf
R}[\epsilon^{-1},[\epsilon]] \ee for some positive integer $k$,
then
\be
R_{MS}(v)=\sum_{i:=0}^{k-1} c_{-k+i}\epsilon^{-k+i}.
\ee
Clearly, $R_{MS}[xy]\not= R_{MS}[x]R_{MS}[y]$.
Nevertheless, $R_{MS}$ fulfils the multiplicativity constraints (m.c.'s)
which we formulate for an arbitrary renormalization map $R$ as
\be
R\left[\prod_{i=1}^r R[x_i]\right]=\prod_{i=1}^r R[x_i]\label{mc1}
\ee
and
\be
R\left[\prod_{i=1}^r (x_i-R[x_i])\right]=0,\label{mc2}
\ee
both valid for some positive integer $r>1$, and arbitrary $x_i\in V$.
Note that, setting $r=2$ and $x_2=1$, the first constraint implies
$R[R[x_1]]=R[x_1]$ and hence $R[x_1-R[x_1]]=0$.
The second constraint establishes the same property also for products of such
differences.

Those constraints can be concluded from a single
condition:\footnote{I thank Christian Brouder for pointing my
attention to this fact.}
\be
R[xy]-R[R[x]y]-R[xR[y]]+R[x]R[y]=0,\label{brouder} \ee which
implies the m.c.'s.


We call these constraints the multiplicativity constraints, as one can show
that for maps $R$ in accordance with these constraints one has\\
{\bf Prop.2:}
\be
S_R\left[\prod_i \phi(t_i)\right]=\prod_i S_R[\phi(t_i)],\;\forall
\phi\in{\cal H}\otimes V.\label{hom} \ee We can prove this
statement under fairly general circumstances:\footnote{This result
has far reaching consequences showing the conceptual significance
of renormalization by its relation to the Riemann--Hilbert problem
\cite{ad3}.} Let ${\cal H}$ be a commutative, graded Hopf algebra
with coproduct $\Delta$, antipode $S$, multiplication $m$, unit
$e$ and counit $\bar{e}$, over some number field ${\cal F}$, and
such that the subalgebra ${\cal H}_0$ of elements of degree zero
is reduced to scalars, so that ${\cal H}_0$ is the kernel of
$(id-E\circ\bar{e})$. Let $E$ be the standard inclusion of ${\cal
F}$ in ${\cal H}$, $E: {\cal F}\to {\cal F}e\in{\cal H}$. Note
that $S^2=id$.


Let a representation $\phi:{\cal H}\to V$ be given. This includes
the case $\phi=id$, $V={\cal H}$ itself. In the following, we
discuss only this case, the changes necessary for the general case
are obvious and demand only the insertion of the map $\phi: {\cal
H}\to V$ at appropriate places. Also, the prove goes through  for
the non-commutative case, delivering the expected homomorphism
property $S_R(XY)=S_R(Y)S_R(X)$.


Let then $R$ be a map $R:{\cal H}\to {\cal H}$ which fulfils
the m.c.'s (\ref{mc1},\ref{mc2}). It need not be an algebra endomorphism,
$R(xy)\not=R(x)R(y)$.
But we demand $R(e)=e$.
Define
$
P_2:{\cal H}\otimes{\cal H}\to{\cal H}\otimes{\cal H}
$
by
\be
P_2=(id-E\circ\bar{e})\otimes (id-E\circ\bar{e}),
\ee
so that we can write
\be
\Delta(X)=P_2(\Delta(X))+e\otimes X+X\otimes e-[E\circ\bar{e}\otimes
E\circ\bar{e}]\Delta(X).\label{cop2}
\ee
Then, the antipode can be written in the form
\be
S(X)=-X-m[(S\otimes id)P_2\Delta(X)]+2e\bar{e}(X), \ee using
$m[(S\otimes id)\Delta(X)]=E\circ\bar{e}(X)$ and (\ref{cop2}). For
any map $R$ as above, we define
\be
S_R(X)=-R[X+m[(S_R\otimes id)P_2\Delta(X)]]+2e\bar{e}(X). \ee This
definition works recursively, as $S_R$ on the rhs is applied to
elements of lower degree than $X$. We also set $S_R(e)=e$. We now
prove Prop.(2):\\ It suffices to prove the assertion for elements
which fulfil $\bar{e}(X)=0$. Otherwise, one decomposes
$X=(X-E\circ\bar{e}(X))+E\circ\bar{e}(X)$.


So let $X,Y$ be two elements which are annihilated by the counit.
Then,
\be
S_R(XY)=-R[XY]-R[U], \ee where $U=m[(S_R\otimes
id)P_2\Delta(XY)]$, hence \bea U & = &
S_R[X]Y+S_R[Y]X+S_R[XY^\prime]Y^{\prime\prime}
+S_R[Y^\prime]XY^{\prime\prime} +S_R[X^\prime Y
]X^{\prime\prime}\nonumber\\
 & & +S_R[X^\prime]X^{\prime\prime}Y
+S_R[X^\prime Y^\prime]X^{\prime\prime}Y^{\prime\prime},
\eea abbreviating $P_2(\Delta(X))=X^\prime\otimes X^{\prime\prime}$,
(omitting the summation sign). Now, as ${\cal H}$ is
a graded Hopf algebra by assumption, we can proceed by induction
on that grading.
For the start of the induction we take the grade
$n=1$, where the assertion is  immediate.

Assume that the assertion holds for products $XY$ of grade $n$, we
prove that it holds for grade $n+1$.


Thus, by assumption we can write $U$ as
\bea
U & = & S_R[X]Y+S_R[Y]X+S_R[X]S_R[Y^\prime]Y^{\prime\prime}
+S_R[Y^\prime]XY^{\prime\prime}
+S_R[Y]S_R[X^\prime]X^{\prime\prime}\nonumber\\
 & & +S_R[X^\prime]X^{\prime\prime}Y
+S_R[X^\prime]S_R[Y^\prime]X^{\prime\prime}Y^{\prime\prime}. \eea
Now, we use the fact that $R$ fulfills the multiplicativity
constraints to write \bea R[XY] & = &
R[X]R[Y]+R[R[X]\tilde{Y}]+R[R[Y]\tilde{X}]\label{mul1}\\ R[U_XU_Y]
& = &
R[U_X]R[U_Y]+R[R[U_X]\tilde{U_Y}]+R[R[U_Y]\tilde{U_X}]\label{mul2}
\eea where $\tilde{Z}:=Z-R[Z], \forall Z\in{\cal H}$ and
\be
U_X:=S_R(X^\prime)X^{\prime\prime}=m[(S_R\otimes id)P_2\Delta(X)]
\ee and similarly for $U_Y$. Now use $S_R(X)=-R[X]-R[U_X]$ which
enables us to completely decompose $U$ and $R[XY]$ in terms of
$R[X],R[Y],R[U_X],R[U_Y]$. Using (\ref{mul1},\ref{mul2}) one finds
\be
-R[XY]-R[U]=(R[X]+R[U_X])(R[Y]+R[U_Y])=S_R(X)S_R(Y),
\ee
as desired.~$\Box$\\



Now, renormalization maps all serve one and the same purpose:
to eliminate the undesired divergences in the theory.
Typically, they can be considered as transformations which do not
alter the behaviour at large internal scales (internal refererring
here to scales which are to be integrated out) such that
one can establish well-defined ratios like the functions $\Gamma_{a,b}(t)$ defined
before. In such ratios
the dependence on large internal scales drops out
and hence we find finite results for renormalized
Green functions.
General renormalization maps $R$  useful in applications
in QFT can be simply considered as maps which fulfil
the multiplicativity constraints, but {\em not}
necessarily the algebra
homomorphism property $R[xy]=R[x]R[y]$.



Realizing that one can label contributions to Green functions
of pQFT by decorated rooted trees in the same manner as we did so far with functions
$G_{b,\infty}^t$, we consider quite generally maps $\phi$ from
${\cal H}$ to some appropriate space $V$.


Let us then introduce, for any such $\phi:{\cal H}\to V$, from
which we demand nothing more than the algebra homomorphism
property $\phi(ab)=\phi(a)\phi(b)$,
\be
\phi_R:{\cal H}\to V,\;\phi_R(t)=S_R[\phi](S(t)) \ee where $R$ is
any map which fulfils (\ref{brouder}) and hence the
multiplicativity constraints (\ref{mc1},\ref{mc2}). $S_R:{\cal
H}\to V$ is still defined as
\be
S_R(\phi)=-R[\phi+m((S_R\otimes id)(\phi\otimes\phi)P_2\Delta].
\ee
Then, as before, let ${\cal R}$ be the corresponding map
\be
{\cal R}: {\cal H}^\star\otimes V\to {\cal H}^\star\otimes V,
\;{\cal R}(\phi)=\phi_R=S_R[\phi\circ S].
\ee
All the maps $\phi_R$ are algebra homomorphisms, due to (\ref{hom}).


Then, we define $\Gamma_{R,\phi}:{\cal H}\to V$ by
\be
\Gamma_{R,\phi}(t)=
m[(\phi_{R}
\otimes
\phi)
(S\otimes id)\Delta(t)].
\ee
For example, setting $\phi=\phi_b$, $R=R_a$ we recover the previous definition.
We clearly have
\be
\Gamma_{R,\phi}(t)=[\phi_R\circ S\star \phi](t).
\ee
Note that  $\Gamma_{R,\phi}$ calculates the ratio of the
two representations $\phi_R,\phi$ with respect to the convolution
product, forming the ratio with the help of the antipode,
as it should be.
In general, one can define, for any two representations
$\phi_u,\phi_v\in{\cal H}^\star\otimes V$,
\be
\Gamma_{u,v}(t)=[\phi_u\circ S\star\phi_v](t).
\ee
Then, the generalized form of Chen's Lemma takes the form
\be
\Gamma_{u,v}(t)=\left[\Gamma_{u,s}\star\Gamma_{s,v}\right](t),
\ee
where $u,s,v$ are labels indexing  different representations
$\phi_u,\phi_v,\phi_s:{\cal H}\to V$. $\phi_{R_a},\phi_b$ were
examples, as are $\phi_{R}$, $\phi$.


Let us show that ${\cal R}\circ{\cal R}(\phi)={\cal R}(\phi)$,
\beas
{\cal R}\circ{\cal R}(\phi) & = & S_R(S_R(\phi))\\
 & = & -R[S_R(\phi)+m((S_R\otimes id)(S_R(\phi)\otimes S_R(\phi))P_2\Delta]\\
 & = & -R[S_R(\phi)+m((S_R\otimes id)(id\otimes S_R)(\phi\otimes \phi)(S\otimes id)P_2\Delta)]\\
 & = &  -R[S_R(\phi)+S_R[m((\phi\otimes \phi)(S\otimes id)P_2\Delta)]]\\
 & = & R[S_R[-\phi-m((\phi\otimes \phi)(S\otimes id)P_2\Delta)]]\\
 & = & R[S_R[\phi\circ S]]=S_R(\phi\circ S)={\cal R}(\phi).
\eeas This proof works by induction on the number of vertices.
From the second to the third line we used the assertion for lesser
than $n$ vertices, by employing $S_R\circ S_R(\phi)={\cal
R}\circ{\cal R}(\phi)={\cal R}(\phi)=S_R(\phi\circ S)$. In the
last line we utilized that $S_R$ maps to the range of $R$ and
$R^2=R$.

Hence, renormalization maps $R$ which fulfill the multiplicativity
constraints fulfill ${\cal R}^2(\phi)={\cal
  R}(\phi)$.
Further, if we have a renormalization scheme $R^\prime$ and
representations $\phi^\prime,\phi\in {\cal H}^\star\otimes V$ such
that $S_{R^\prime}(\phi^\prime)=\phi\circ S$ (that is, the natural
generalization of (\ref{cha}) holds), then Lemma 1 holds in the
form
\be
\Gamma_{R,\phi^\prime}=
\Gamma_{R,\phi}\star\Gamma_{R^\prime,\phi^\prime},\label{lll} \ee
where $R$ can be any renormalization scheme. Here,
$\phi^\prime_{R^\prime}:=S_{R^\prime}(\phi^\prime\circ S)$, and
the condition $S_{R^\prime}(\phi^\prime)=\phi\circ S$ essentially
guarantees that we renormalize the second term on the rhs of
(\ref{lll}) at a renormalization point at which the unrenormalized
functions $\phi$ in the first term on the rhs are evaluated.
Hence, the lhs is a concatenation of two renormalized Green
function on the rhs, the first evaluating bare Green functions at
parameters which we use for renormalization of the second, which
shifts the bare function from $\phi$ to $\phi^\prime$. The
infinitesimal version of this reparametrization can be regarded as
the generator of the flow of the renormalization group.


Indeed, in (\ref{lll}) we see that the rhs depends on the
intermediate representation $\phi$, which does not appear on the
lhs. If we regard $\phi$ as charaterized by an appropriate (set
of) parameter(s) $b$, and $\phi^\prime$ characterized by a (set
of) parameter(s) $b^\prime$, we can apply a differentiation with
respect to $b$ to find
\be
0=\frac{d}{db}
\left[
\Gamma_{R,\phi}\star\Gamma_{R^\prime,\phi^\prime}\right]
 =
\left(\frac{\partial}{\partial b}
\Gamma_{R,\phi}\right)\star
\Gamma_{R^\prime,\phi^\prime}
+
\Gamma_{R,\phi}\star
\left(\frac{\partial}{\partial b}
\Gamma_{R^\prime,\phi^\prime}\right),\label{rge}
\ee
which is a proto-type renormalization group equation.
It expresses as a differential equation the independence
of an intermediate scale. Note that in the limit
$b\to b^\prime$ the finite ratio
$\Gamma_{R^\prime,\phi^\prime}$ becomes an infinitesimal
quantity. Note further that the dependence on $b$
of the second term on the rhs is given by the fact that $S_{R^\prime}(\phi^\prime)=
\phi\circ S$.
The exercise to cast the renormalization
group explicitly in this language is a  purely  notational one,
taking into account the dependence on parameters like charges,
masses etc and hence establishing a coupled systems of such equations,
which we postpone to future work. Various viewpoints about
renormalization group equations ranging from standard BPHZ approaches
to the Wilson viewpoint can be obtained from (\ref{lll},\ref{rge})
depending on which parameters for an intermediate scale one chooses,
with dimensionful parameters of bare Green functions or physical cut-offs
being some obvious choices.
\subsection{Cohomological properties of renormalization}
Let us consider the following problem.
Given is a perturbative QFT, defined by Feynman rules.
This defines a series in graphs graded by the number of vertices.
The graphs translate to unique analytical expressions,
which decompose into Feynman integrands and integrations,
determined by the closed loops
in the graph. Powercounting establish a well-defined subset of superficially
divergent subgraphs, and eventually, we realize that each of these graphs
represents an element of ${\cal H}$ \cite{hopf,CK,DiDa}.


The analytic expressions are parametrized by external momenta and
masses, which can be regarded as complex parameters generalizing
the external scale $b$ of the iterated integral. For a given
Feynman graph $\Gamma$, let for now $b_\Gamma$ represent this set
of parameters. The integration is over internal loop momenta along
propagators (edges)  (or over internal vertices, in $x$-space) and
diverges (in momentum space) when the internal momenta get large.
Let us specify a renormalization scheme by saying that for any
graph $\Gamma$ we have defined a set of conditions on the
parameters $b_\Gamma$, for example conditions that the square of
external momenta equals some mass square. Let $\mu_\Gamma$ be the
set of parameters $b_\Gamma$ specified in accordance with these
conditions. These parameters are provided by the integrand
constructed according to the Feynman rules.\footnote{There is an
analogous story in x-space to be developped elsewhere.}


The renormalized Green function established by this set-up can be calculated
as
\be
\Gamma_{R_{\mu_\Gamma},\phi_{b_\Gamma}}(t_\Gamma)=
[S_{R_{\mu_\Gamma}}(\phi_{b_\Gamma})\star \phi_{b_\Gamma}](t_\Gamma),
\ee
where $t_\Gamma\in{\cal H}$ is obtained from $\Gamma$, and
$\phi_{b_\Gamma}$
maps
it to an analytic expressions according to the Feynman rules
and all other notation is self-evident.


This can be written in the form, in an obvious shorthand notation,
\be
\Gamma_{R,\phi}=
m\circ({\cal R}_\mu\otimes id)(\phi_b\otimes \phi_b)\circ (S\otimes id)\Delta,
\ee
which shows that
 we fail by the deviation of ${\cal R}$ from the identity (in ${\cal H}^\star\otimes V$)
to get a trivial result.
The interesting operator in the above is clearly ${\cal R}_\mu\otimes id$.


Let us now concatenate the renormalization step $n$ times. Hence,
we assume that we have given a renormalized Green function
$\Gamma_R,\phi_{b_0}$ as above, where the notation stresses that
we use a set of parameters $b_{0\Gamma}$ for the bare function,
and some arbitrary renormalization $R$.

Let us now vary these external parameters through $n$ steps until
they reach a bare Green function $\phi_{b_n}$, which uses other
values for external parameters. In each step, we will use a
renormalized Green function which subtracts $\phi_{b_i}(t_\Gamma)$
at $\mu_\Gamma=b_{i-1,\Gamma}$, thus
$\phi_{{b_i}R_{b_{i-1}}}=\phi_{b_{i-1}}$. Hence, we achieve a
concatenation of renormalizations where each step fullfils
(\ref{lll}).

Thus consider a sequence of renormalizations sending
$\phi_{b_0}\to\phi_{b_n}$ by an intermediate sequence of
renormalizations $b_0\to b_1\to\ldots\to b_n$, using \bea
\phi_{b_{i-1}}\star\Gamma_{R_{b_{i-1}},\phi_{b_i}} & = &
\phi_{b_{i-1}}\star{\cal R}_{b_{i-1}}(\phi_{b_i}\circ
S)\star\phi_{b_i} \nonumber\\
 & = &
\phi_{b_{i-1}}\star \phi_{b_{i-1}}\circ S\star\phi_{b_i}
\nonumber\\
 & = & \phi_{b_{i-1}}[id\star S]\star \phi_{b_i}\nonumber\\
 & = & \phi_{b_{i-1}}[\bar{e}]\star\phi_{b_i}\nonumber\\
 & = & \phi_{b_i}.
\eea


Let us introduce
\be
d_R:{\cal H}^\star\otimes V\to({\cal H}^\star\otimes V)^{\otimes 2}
\ee
by
\be
d_R(\phi_{b_i})={\cal R}_{b_{i-1}}(\phi_{b_i})\otimes
\phi_{b_i},
\ee
so that we obtain
\bea
\phi_{b_n} & = & M[[\phi_{b_1}\otimes d_{R_{b_1}}(\phi_{b_2})\otimes
d_{R_{b_2}}(\phi_{b_3})\otimes\ldots
\otimes d_{R_{b_{n-1}}}(\phi_{b_n})]\nonumber\\
 & & \circ(id\otimes[S\otimes
id]^{\otimes(n-1)}
\Delta^{2n-1})],
\eea
where $M$ is a concatenation of $2n$ multiplication maps,
$M:V^{\otimes 2n+1}\to V$, and $\Delta^{2n+1}: {\cal H}\to {\cal H}^{\otimes
2n+1}$ is the obvious map in ${\cal H}$
sending ${\cal H}\to {\cal H}^{\otimes 2n+1}$ using the coproduct,
unique due to coassociativity of the latter.


This motivates to define $d_R$ for elements
$({\cal H}^\star\otimes V)^{\otimes n}$.
\be
d_R(\phi_1\otimes \phi_2\otimes\ldots\otimes \phi_n)= \sum_{i=1}^n
(-1)^{i+1}\phi_1\otimes \ldots \otimes
d(\phi_i)\otimes\ldots\otimes\phi_n, \ee where it is understood
that $d(\phi_i)={\cal R}_{i-1}(\phi_i)\otimes \phi_i$, $i>1$ and
$d(\phi_1)=d_R(\phi_1)={\cal R}(\phi_1)\otimes\phi_1$. Hence, at
the $i$-th entry ($i>1$), ${\cal R}$ replaces $\phi_i$ by the
element ${\cal
R}_{i-1}(\phi)\otimes\phi_i=\phi_{i-1}\otimes\phi_i$. At the first
entry, we obtain ${\cal R}(\phi_1)$. It is convenient to introduce
$\phi_0:={\cal R}(\phi_1)$, with ${\cal R}(\phi_0)=\phi_0$.


Then, one immediately checks, using ${\cal R}^2={\cal R}$,
\be
d_R^2=0,
\ee
for example:
\be
d_R(d_R(\phi_1))=d_R(\phi_0\otimes\phi_1)=\phi_0\otimes
\phi_0\otimes \phi_1-\phi_0\otimes {\cal R_0}(\phi_1)\otimes
\phi_1=0. \ee In this language, we have
\be
\phi_{b_n}=M[[\phi_{b_1}\star d(\phi_2)(S\otimes id)
\star\ldots\star d(\phi_n)
(S\otimes id)]\Delta^{2n+1}]
\ee
and for the renormalized Green function
\be
d_R(\phi_n)=M[d(\phi_1)(S\otimes id)\star(d\phi_2)(S\otimes id)
\star\ldots\star(d\phi_n)(S\otimes id)\Delta^{2n+2}].
\ee
In short, taking into account that the actions of $M,\Delta^{\ldots},(S\otimes id)$
are obvious:
\be
\phi_n=\phi_1d\phi_2\ldots d\phi_n
\ee
and
\be
d\phi_n=d\phi_1\ldots d\phi_n.
\ee
We recommend that the reader tries these formula out on several simple
examples and marvels at their obvious cohomological
relevance especially in comparison with \cite{ACbook}.


We realize that the change of scales, so typically a step in
the whole of physics, naturally carries cohomological structure
which gives hope to be able to cast locality in a well-defined
mathematical framework in the future.



Much more can and should be said about these aspects. Here, we have to refer the
reader to future work \cite{CKnew}.


\subsection{Tree-indexed scales and the equivalence of schemes}
Let us come back to Chen's Lemma and to iterated integrals.
There are further generalizations
lying ahead. So far, we used iterated integrals as
quantities which are naturally indexed by rooted trees, and to which the
previous considerations apply.
The rooted trees determined how
the various differential forms $f_i(x)dx$ are combined under the
(indefinite) integral operator, but outer boundaries where kept constant
throughout.
\subsubsection{Tree indexed scales}
A generalization which turns out to be quite useful in practice
is to let even these  boundaries be indexed by decorated rooted
trees. Hence, we redefine
\be
-G_{b,\infty}(t)=\int_{b_t}^\infty f_{i_n}(x)\prod
G_{x,\infty}^{t^\prime}dx \ee which is defined as before, only
that we now label the lower outer boundary by a decorated rooted
tree $t$. We  understand that the map $R_a$ maps lower boundaries
$b_t$ to $a_t$, and that the coproduct action extends to this
label. Similar considerations apply to full-fledged Green
functions of pQFT, where one can utilize the presence of scale
dependent parameters to make them tree-dependent in the same
manner. This idea will be pursued in the next section, and in
\cite{DKnew}.



If $t_2(f_1,f_2)$ is the decorated rooted tree of Fig.(\ref{f1}) with coproduct
\be
\Delta(t_2(f_1,f_2))=t_2(f_1,f_2) \otimes e+e\otimes
t_2(f_1,f_2)+t_1(f_1)\otimes t_1(f_2) \ee we formally obtain
(abbreviating
$t_2(f_1,f_2)=t_{2_{12}},t_1(f_1)=t_{1_1},t_1(f_2)=t_{1_2}$),
\be
S_{R_a}(\phi_{b})(t_2(f_1,f_2))= \left[
-\int_{a_{t_{2_{12}}}}^\infty
\int_{x}^\infty+\int_{a_{t_{1_2}}}^\infty\int_{t_{1_1}}^\infty\right]f_2(x)f_1(y)dydx.
\ee and
\be
\Gamma_{a,b}(t_2(f_1,f_2))=\left[\int_{b_{t_{2_{12}}}}^\infty
\int_x^\infty- \int_{b_{t_{1_2}}}^\infty
\int_{a_{t_{1-1}}}^\infty+ \int_{a_{t_{1_2}}}^\infty
\int_{a_{t_{1_1}}}^\infty-\int_{a_{t_{2_{12}}}}^\infty\int_x^\infty\right]f_2(x)f_1(y)dydx.
\ee In this notation, $a,b$ are to be regarded as representing
actually a whole set of constants $a_t,b_t$, parametrizing the
relevant scales for the decorated tree considered.


Still, (\ref{gchen}) applies and describes what happens if we change the renormalization
point. $\Gamma_{a,b}=\Gamma_{a,s}\star\Gamma_{s,b}$
now becomes
\beas
 & & \left[
\int_{b_{t_{2_{12}}}}^\infty \int_x^\infty-
\int_{b_{t_{1_2}}}^\infty \int_{a_{t_{1_1}}}^\infty+
\int_{a_{t_{1_2}}}^\infty
\int_{a_{t_{1_1}}}^\infty-\int_{a_{t_{2_{12}}}}^\infty
\int_x^\infty\right]f_2(x)f_1(y)dydx\\
 & = &
\left[ \int_{s_{t_{2_{12}}}}^\infty \int_x^\infty-
\int_{s_{t_{1_2}}}^\infty \int_{a_{t_{1_1}}}^\infty+
\int_{a_{t_{1_2}}}^\infty
\int_{a_{t_{1_1}}}^\infty-\int_{a_{t_{2_{12}}}}^\infty\int_x^\infty\right]
f_2(x)f_1(y)dydx\\
 &  &
+\left[\int_{b_{t_{2_{12}}}}^\infty \int_x^\infty-
\int_{b_{t_{1_2}}}^\infty \int_{s_{t_{1_1}}}^\infty+
\int_{s_{t_{1_2}}}^\infty \int_{s_{t_{1_1}}}^\infty
-\int_{s_{t_{2_{12}}}}^\infty\int_x^\infty\right]f_2(x)f_1(y)dydx\\
 & &
+\left[\int_{s_{t_{1_1}}}^\infty
-\int_{a_{t_{1_1}}}^\infty\right]f_1(y)dy
\left[\int_{b_{t_{1_2}}}^\infty -\int_{s_{t_{1_2}}}^\infty\right]
f_2(x)dx, \eeas which is evidently true, as the reader should
check. It is instructive to see the Lemma (1) in action for this
simple example.


Finiteness of $\Gamma_{a,b}$ now imposes conditions on the
tree-indexed parameters, a fact which we  will utilize in the next section.


\subsubsection{Equivalence of schemes}
Conceptually, the presence of tree-indexed parameters allows to describe
different renormalization schemes on a similar footing.
The idea is the following.
Let us compare for example
a BPHZ on-shell scheme in comparison
with minimal subtracted dimensional renormalization.
In the former case, one effectively
subtracts at the level of integrands. Hence, one has an integrand
which gives rise to a non-existent measure with respect to the
loop integrations. The integrand is parametrized by several
constants (masses, external momenta),
which essentially play the role of the boundaries in our
iterated integrals. A renormalization scheme in the BPHZ spirit
would map, upon applying the antipode
$S_{R_{BPHZ}}$, such an integrand to another one, for which these parameters
fulfill certain conditions (on-shell, for example).
The structure of $S_{R_{BPHZ}}$ achieves that these counterterms are
local \cite{hopf,overl,CK}.


Then, the map $\Gamma_{R_{BPHZ},\phi}(t)$ so-constructed
delivers a subtracted integrand which actually establishes a well-defined
measure with respect to all loop integrations. Here, $t$ is the decorated
rooted tree assigned to the integrand according to powercounting
\cite{hopf,CK,overl,DiDa}.
Hence, BPHZ-type schemes avoid the use of regularization altogether.


On the other hand, in dimensional renormalization using minimal
subtraction (MS scheme), one introduces regularization and evaluates the bare
Green-functions first, obtaining Laurent series in the
regularization parameters. The antipode achieves a subtraction of
these poles which respects locality, and similarly one constructs
the MS-renormalized $\Gamma_{R_{MS},\phi}(t)$ using $S_{R_{MS}}$.


In the next section we will use the idea to
have tree-indexed scales to show that we can regard
a MS-scheme as a BPHZ type scheme on the expense of having
to introduce tree-dependent scales.
\section{Applications}
To keep the amount of notation simple, we will consider
representations of ${\cal H}$ defined as follows. Assume that we
are given a set of functions $B_k(x)$ which are Laurent series in
$x$ with a first-order pole. Using vertex weights and the
corresponding notation as defined in the beginning of
section five
below we can define a function
\be
G_z(t;x)=\prod_{v\in T^{[0]}} B_{w(v)}(x)z^{-nx}=B_t(x)z^{-nx},
\ee
where $n$ is the number of vertices of $t$ and
$z$ is to be regarded as the scale parametrizing the
representation and $x$ is the regularization
parameter. We also wrote
$\prod_{v\in T^{[0]}} B_{w(v)}(x)=B_t(x)$.
Quite a number of interesting
applications can be brought to this form \cite{DiDa,DKnew}.
Typically, whenever we iterate a Feynman diagram in terms of itself
as described by a rooted tree, one finds such representations.
Many examples are given in \cite{DiDa}.
The  case of general Feynman diagrams is obtained by finding a proper
notation for the case of different decorations,
and by taking into account a proper form-factor decomposition.
We refer the reader to \cite{DKnew}
for further applications, extending to such cases.


Hence, in the notation of the previous section,
we set $\phi_z(t)=G_z(t;x)$.
Then, we define the MS renormalization scheme by setting
\be
R_{MS}\circ \phi_z=<\phi_1>,
\ee
where angle brackets denote projection on the pole part of the
Laurent series in
$x$ inside the brackets. Let then the counterterm defined
by $S_{R_{MS}}(\phi_z)$ and the renormalized Green function
by $\Gamma_{MS}(\phi_z)(t)=[S_{R_{MS}}(\phi_z)\star \phi_z](t)$,
as usual.
The reader should have no difficulties confirming
that for  $t_2$,  the rooted tree with two vertices,
\be
S_{R_{MS}}(\phi_z)(t_2)=-<B_2B_1>+<<B_1>B_1>,
\ee
and
\be
\Gamma_{R_{MS},\phi_z}(t_2)=B_2 B_1z^{-2x}-<B_1>B_1z^{-x}-<B_2B_1>+<<B_1>B_1>.
\ee
Let us compare such an approach with an on-shell approach. Using the same bare
functions, we define the on-shell renormalization map $R_\mu$ as
\be
R_{\mu}\circ\phi_z=\phi_\mu.
\ee
If the $G_z(t;x)$ are provided by integrals whose integration
is regularized by $x$ where $z$ is a parameter of the integrand, then
the renormalization map just sets this external parameter to the value $\mu$.


$\Gamma_{R_\mu,\phi_z}(t)=[S_{R_\mu}[\phi_z]\star\phi_z](t)$
is a function which has a subtracted integrand such that typically
the limit $x\to 0$ exists at the level of the integrand. It becomes
a Taylor series in $\log(z/\mu)$.


Let us now cast the MS renormalized Green function in this form
on the expense of introducing tree-dependent scales $\mu_t$.


Hence, we redefine $R_\mu(\phi)(t)=\phi_{\mu_t}(t)$,
and get, still spelling out the example $t=t_2$,
\be
S_{R_{\mu}}(\phi_z)(t_2)=
-B_2B_1\mu_{t_2}^{-2x}+B_1B_1\mu_{t_1}^{-2x}.
\ee
We remind ourselves that $S(t_2)=-t_2+t_1 t_1$.


It is easy to work out the general case. To see this, we look at
the trees with up to three vertices.
The antipode $Z_t:=S_{R_{MS}}(\phi_z)(t)$ in MS reads
\begin{eqnarray}
Z_{t_1} & = & -<B_1>,\\
Z_{t_2} & = & -<B_2B_1>+<<B_1>B_1>,\\
Z_{t_{3_1}} & = & -<B_3B_2B_1>+<<B_1>B_2B_1>+<<B_2B_1>B_1>\nonumber\\
 & & -<<<B_1>B_1>B_1>,\\
Z_{t_{3_2}} & = & -<B_3B_1^2>+2<<B_1>B_2B_1>\nonumber\\
 & & -<<<B_1>B_1>B_1>.
\end{eqnarray}
In general, one finds
\be
Z_t=
\sum_{\mbox{\tiny full cuts $C$ of $t$}}(-1)^{n_C}<\left[\prod_i <B_{t_i}>\right]B_{t_R}>.
\ee
The antipode $Z_t^\mu$ in a subtraction scheme using tree-indexed parameter
sets $\mu_t$ reads
\be
Z_t^\mu=
\sum_{\mbox{\tiny full cuts $C$ of $t$}}(-1)^{n_C}
\left[\prod_i B_{t_i} \mu_{t_i}^{-\#(t_i)
  x}\right]B_{t_R}\mu_{t_R}^{-\#(t_R) x}.
\ee
We remind the reader that $\#(t)$ equals the number of vertices of $t$.


Equating $Z_t=Z_t^\mu$ determines $\mu_t$ recursively
\be
\mu_t=
\exp\left[\left(\frac{-1}{x t^!}\right)
\log(B^\prime_t/B_t)\right],
\label{mup}
\ee
where
\be
B^\prime:= \sum_{\mbox{\tiny full cuts $C$ of $t$}}(-1)^{n_C}
\left[\prod_i S_{R_{MS}}(\phi_z)(t_i)\right]
S_{R_{MS}}(\phi_z)(t_R).\label{bprime} \ee


One also confirms that now
\be
\Gamma_{R_{MS},\phi_z}(t)=\Gamma_{R_\mu,\phi_z}(t)
\ee
holds.
Note that we can discard the use of a regularization scheme in
$\Gamma_{R_{MS},\phi_z}$ as we observe that the scales $\mu_t$ are functions
which exist in the limit $x\to 0$, which one confirms
by using $B^\prime_t/B_t=1+{\cal O}(x)$ in (\ref{mup}).


Conceptually, this eliminates
any difference between a BPHZ type scheme and regularization followed
by a minimal subtraction. Each divergent subgraph can be subtracted at
its own scale, such that a subtraction with such tree-dependent sets
of parameters equals the result of the use of a MS scheme.
Actually, there remains an argument in favour of minimal
subtraction: it incorporates from the beginning the wisdom
that it is only logarithmic divergence which counts.
Any integrand providing a different degree of divergence
can be cast in the form of a log divergent integrand, (multiplied
by a polynomial in external parameters) plus scale-independent terms.\footnote{An
easy example: $\int_1^\infty \frac{xdx}{x+c}=-c\int_1^\infty \frac{dx}{x+c}+\int_1^\infty
dx$.}
The latter do not contribute anyhow after renormalization, and
are economically eliminated  in dimensional renormalization from
the beginning.


We know that $B_t$ is of order
${\cal O}(1/x^{\#(t)})$, as is $B^\prime(t)$.
One immediately proves that the difference is of lower order
\be
[B_t-B^\prime_t]\sim {\cal O}(1/x^{\#(t)-1}).
\ee
This is a direct consequence of the fact that the antipode $S$ of ${\cal H}$
fulfills $S^2=1$.
Indeed, to leading order in $1/x$ we have
\begin{eqnarray}
B^\prime_t & = & \sum_{\mbox{\tiny full cuts $C$ of $t$}}(-1)^{n_C}<[\prod_i
<Z_{t_i}>Z_{t_R}>\\
 & = & S_{R_{MS}}[Z_t]\\
  & = & Z_{S(t)}=\phi_z(S^2(t))=B_t,
\end{eqnarray}
where we used that to leading order $<<A>B>=<A><B>$ for arbitrary
expressions $A,B$ so that the first line to leading order agrees
with (\ref{bprime}), and we used (\ref{cha}). As these properties
are true not only for minimal subtraction but for any
renormalization scheme, we conclude that in any renormalization
scheme the leading pole part is the same, a property well-known to
the practitioner. It is nice to see it traced back to the fact
that $S^2=1$.


Once more, it is instructive to write down the first couple of cases for $B^\prime_t$.
\begin{eqnarray}
B_{t_1}^\prime & = & <B_{t_1}>\\
B_{t_2}^\prime & = & <B_{t_2}>-<<B_{t_1}>B_{t_1}>+<B_{t_1}><B_{t_1}>\\
B_{t_{3_1}}^\prime & = & <B_{t_{3_1}}>-<<B_{t_1}>B_{t_2}>-<<B_{t_2}>B_{t_1}>
\nonumber\\
 & &
+<<<B_{t_1}>B_{t_1}>B_{t_1}>
-2<B_{t_1}><B_{t_2}>\nonumber\\
 & &  +2<B_{t_1}><<B_{t_1}>B_{t_1}>-<B_{t_1}><B_{t_1}><B_{t_1}> \\
B_{t_{3_2}}^\prime & = & <B_{t_{3_2}}>-2<<B_{t_1}>B_{t_2}>
\nonumber\\
 & &
+<<B_{t_1}><B_{t_1}>B_{t_1}>
-2<B_{t_1}><B_{t_2}>\nonumber\\
 & &  +2<B_{t_1}><<B_{t_1}>B_{t_1}>-<B_{t_1}><B_{t_1}><B_{t_1}>
\end{eqnarray}
It is not a big surprise to see that a change of renormalization
scheme can produce a lot of junk at subleading orders.



A final remark concerns momentum schemes (schemes which subtract
at specified values of external momenta), hence schemes which
fulfil (\ref{cha}). We know already that the convolution product
holds for arbitrary renormalization schemes. The structure of the
convolution product indicates how we translate a renormalized
Green function, $\Gamma_{R,\phi_b}(t)$, determined by a scheme $R$
and parameter(s) $b$, to $\Gamma_{R,\phi_{b^\prime}}(t)$. Both
utilize the same freely chosen renormalization scheme $R$ giving
an operator $d_R$, $d_R^2=0$. The transition $b\to b^\prime$  uses
the  convolution by reparametrizations of the external
parameter(s) $b$, hence a convolution using momentum schemes. They
thus typically provide the general mediator for renormalized Green
functions, as was exhibited in \cite{DiDa,DKnew}.


\section{Tree factorials and CM weights}
In this section we want to prove some basic results concerning
tree-factorials and Connes Moscovici weights. Both entities are
combinatorial in nature. For simplification, we work in the
undecorated Hopf algebra. Similar identities were derived by
Butcher in his work on numerical integration methods
\cite{But,Brou}. As our derivation is different we still give it
in some detail.


Let $T^{[0]}$ be the set of vertices of the rooted tree $T$. For
any  vertex $v$ of $T$, let $t_v=P^c(T)$, where $c$ is the single
cut which removes the edge incoming to $v$. If $v$ is the root, we
set $t_v=t$. Also, $\#(t_v)$ is the number of vertices of the
monomial $t_v$. Then, we define the tree factorial by
\be
T^!=\prod_{v\in T^{[0]}}\#(t_v)\equiv \prod_{v\in T^{[0]}}w(v),
\ee which also defines the vertex weights $w(v)$. Fig.(\ref{f3})
gives instructive examples. Finally, we set $e^!=1$.
\bookfig{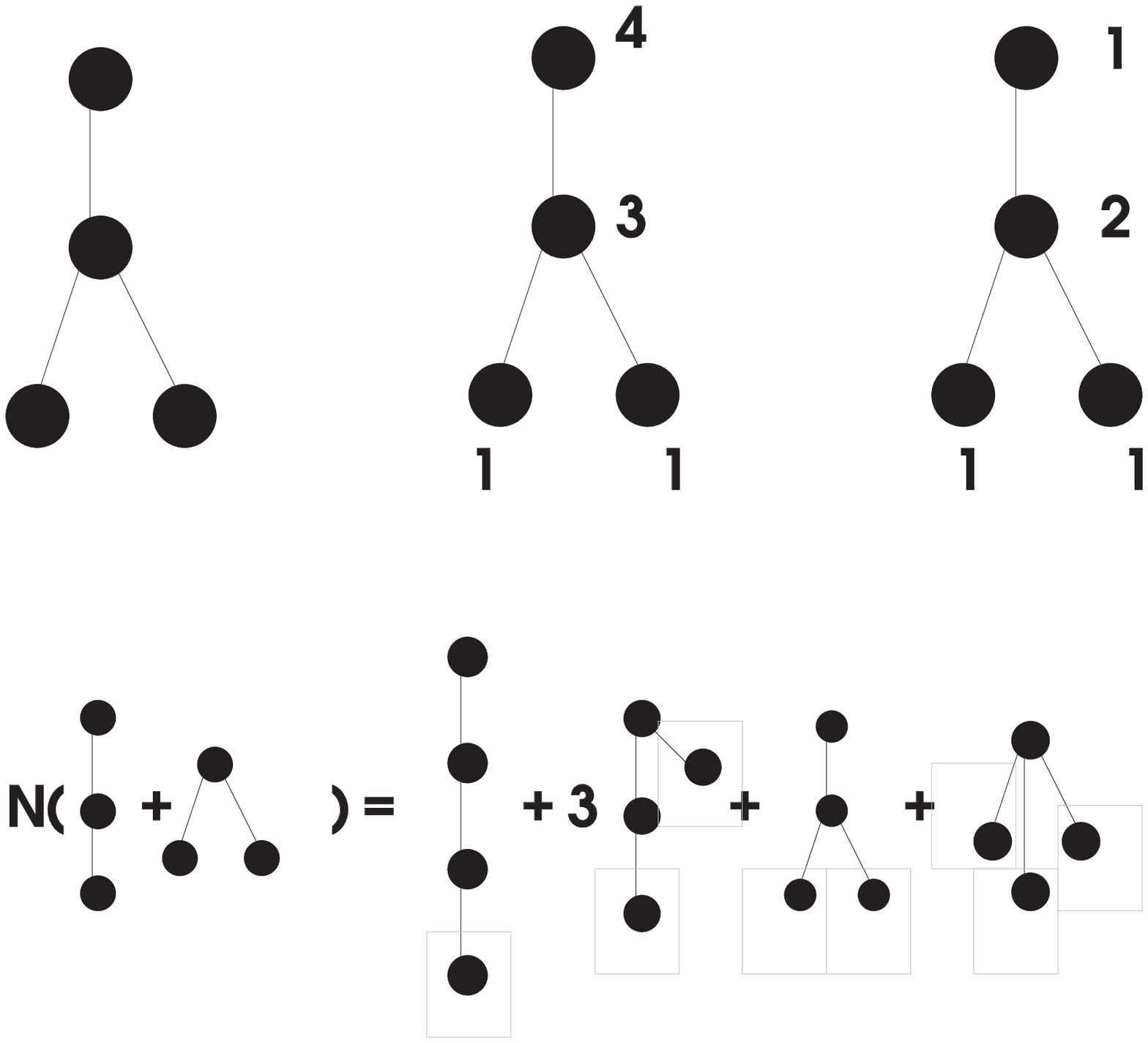}{tf}{f3}{We define vertex weights and tree
factorials, symmetry factors, CM-weights and feet of an
undecorated rooted tree. For the tree $t$ given at the left in the
upper row we indicate the vertex weights at each vertex in the
middle and the vertex symmetries at the right. The second row
considers natural growth of $\delta_3=t_{3_1}+t_{3_2}$. It obtains
on the rhs of the equation four trees with CM-weights $1,3,1,1$.
We also indicate the feet of these four trees.}{6}


The next definition concerns the symmetry factor $S_T$ of a tree $T$.
For any $v\in T^{[0]}$ consider $B_-(t_v)$.
This is a monomial in rooted trees, hence a product of branches
\be
B_-(t_v)=
\prod_{i=1}^{f(v)} t_{v,i}.
\ee
In general, some of these branches can be the same rooted tree,
and hence these products can be written as products $\prod^\prime$
over different rooted
trees with integer powers:
\be
\prod_{i=1}^{f(v)} t_{v,i}=\mbox{$\prod^\prime$}_j [t_{v,j}]^{n_j}.
\ee
We associate to the vertex $v$ its vertex symmetry, built from the factorials
of the multiplicities
 $n_j$ with which a tree $t_{v,j}$ originates from $v$,
\be
S_{T,v}=\mbox{$\prod^\prime$}_j n_j!
\ee
and define
\be
S_T=\prod_{v\in T^{[0]}} S_{T,v}.
\ee
Fig.(\ref{f3}) gives examples for these notions, which were already used
in \cite{DiDa}. In that paper, the symmetry factor $S_T$
was dubbed $\Pi(T)$ and the vertex symmetry of the root $S_{T,r}$
was simply denoted as $\pi(T)$.\footnote{For
a tree $T$ with root vertex $r$ one can consider $S_{T,r}$ to define
what could be dubbed the Moebius function $\mu(T)$ of $T$:
$\mu(T)=0$ iff the branches at $r$ are not square-free (hence if some of the powers
$n_i>1$) and $\mu(T)=1$ if $f(r)$ is even, and $\mu(T)=-1$ if $f(r)$
is odd.}


We now define the Connes Moscovici weights $CM(T)$ for a tree $T$ with $n$ vertices as
\be
CM(T):=<Z_T,\delta_n>,
\ee
where $<\cdot ,\cdot>$ is the pairing of \cite{CK},
$<Z_T,T^\prime>=\delta_{T,T^\prime}$, and $\delta_n=N^n(e)$
is the $n$-fold application of natural growth to $e$, delivering
the generators of the commutative part of the CM-Hopf subalgebra
of ${\cal H}$, obtained by natural growth $N$ applied $n$ times,
cf.~Fig(\ref{f3}).


Two further definitions are useful. The first concerns the feet of
a rooted tree $T$. It is a subset ${\cal F}(T^{[0]})$ of $T^{[0]}$
provided by those vertices which have fertility zero, hence no
outgoing edges. Then, ${\cal F}(T)$ is the set of trees consisting
of all trees which have one foot removed. The cardinality of this
set equals the number of feet of $T$.\footnote{What we call feet
here is often called leaves in the literature on graphs.}


Also, we let ${\cal N}(T)$ be the set of those trees
which are generated from $T$ by natural growth $N(T)$
\be
T^\prime \in {\cal N}(T)\Leftrightarrow <Z_{T^\prime},N(T)>\not= 0,
\ee
and counting multiplicities appropriately.
See Fig.(\ref{f3}) to get acquainted with all these notions.




We want to derive the following three results. First,
\be
\frac{n}{T^!}=\sum_{t\in{\cal F}(T)}\frac{1}{t^!},\label{tf}
\ee
which leads to the second result,
\be
CM(t)=\frac{n!}{t^!S_t},\forall t\in {\cal T}^{[n]},\label{CM} \ee
where the tree-factorials and Connes-Moscovici weights are defined
as in \cite{DiDa}. Here, ${\cal T}^{[n]}$ is the set of trees with
$n$ vertices. Finally,
\be
\sum_{T\in {\cal T}^{[n]}} \frac{CM(T)}{T^!}=\frac{(n-1)!}{2^{(n-1)}}.\label{46}
\ee


We first note that (\ref{tf}) obviously holds for trees
$t=B_+^n(e)$, where it reduces to the familiar
$n/n!=1/(n-1)!$. Hence, the tree factorial is another example
of the replacement {\em integers to rooted trees}.


The next observation is
\be
\frac{n}{t^!}=\prod_{j=1}^k \frac{1}{t_j^!},
\ee
where $B_-(t)=\prod_{j=1}^k t_j$.
This identity is a mere way of writing  the definition of the
tree factorial, using that any tree factorial factorizes $n$ for a
tree with $n$ vertices.


Instead of proving (\ref{tf}), we prove
\be
\frac{
n_1+\ldots+n_k
}{
T_1^!T_2^!\ldots T_k^!
}
=
\sum_{t^\prime\in{\cal F}(B_+(t_1\ldots t_k))}
B_-(t^\prime)^!,\label{p2p} \ee where we define
\be
B_-(t^\prime)^!=\frac{T_i^!}{T_1^!\ldots T_k^!}\frac{1}{(T_i^\prime)^!}
\ee
and $T_i^\prime$ is defined by $B_+(T_1\ldots T_i^\prime\ldots T_k)=t^\prime$.
Thus, we use that the feet must have been attached
to  some branch $T_i$ of $B_+(T_1\ldots T_k)$
which gives $T_i^\prime\in {\cal F}(T_i)$.


Let us first show that (\ref{tf}) is a consequence of (\ref{p2p}).
To see this, it suffices to note that
\be
{\cal F}(B_+(t))=\{B_+(t^\prime)\mid t^\prime\in{\cal F}(t)\},
\ee
as obviously root and feet are different ends of a rooted tree.
Fig.(\ref{f3}) gives an instructive example.
Hence, setting $k=1$ in (\ref{p2p}),
\be
\frac{n}{t^!}=\sum_{t^\prime\in {\cal F}(B_+(t))}B_-(t^\prime)^!=
\sum_{t^\prime\in{\cal F}(t)}\frac{1}{t^{\prime!}}, \ee which
proves (\ref{tf}).~$\Box$\\


It remains to prove (\ref{p2p}). We have, using induction on the
total number of vertices in $T_1\ldots T_n$ and the fact that a
single $T_i$ has lesser vertices than the product,
 \bea \frac{n_1+\ldots
+n_k}{T_1^!\ldots T_k^!} & = &
\sum_{i=1}^k\frac{n_i}{T_i^!}\frac{T_i^!}{T_1^!\ldots T_k^!}\\
 & = & \sum_{i=1}^k\sum_{t_i^\prime\in {\cal F}(T_i)}\frac{1}{(t_i^\prime)^!}
\frac{1}{T_1^!\ldots {\wedge\atop T_i^!}\ldots T_k^!}\\
 & = &
\sum_{t^\prime\in {\cal F}(B_+(T_1\ldots
T_k))}[B_-(t^\prime)]^!.\;\Box \eea In the above, the $\wedge$
means to omit the corresponding tree factorial.



We are now in the position to prove (\ref{CM}).
We assume it holds for trees with $n-1$ vertices
and simply reduce it to (\ref{tf}) to show that it holds for $n$ vertices.
\bea
S_tC\!M(t) & = & \frac{n!}{t^!}\\
 & = & \sum_{t^\prime\in{\cal F}(t)}C\!M(t^\prime)s_{t^\prime}\nonumber\\
 & = & (n-1)!\sum_{t^\prime\in {\cal F}(t)}\frac{S_{t^\prime}}{S_{t^\prime}t^{\prime!}}\\
 & = & (n-1)!\sum_{t^\prime\in {\cal F}(t)}\frac{1}{t^{\prime!}},
\eea
which is the desired result.~$\Box$\\
It remains to prove (\ref{46}).
By the definition of the CM-weights as counting multiplicities
of trees under natural growth we can write
\be
\sum_{t\in {\cal T}^{[n]}}\frac{CM(t)}{t^!}=\sum_{t\prime\in{\cal T}^{[n-1]}}
\frac{CM(t^\prime)}{(t^\prime)^!}
\sum_{t^{\prime\prime}\in {\cal N}(t^\prime)}
\frac{(t^\prime)^!}{(t^{\prime\prime})^!}.
\ee
Hence we must show
\be
\sum_{t^{\prime\prime}\in {\cal N}(t^\prime)}
\frac{(t^\prime)^!}{(t^{\prime\prime})^!}
=\frac{n-1}{2}.\label{br}
\ee
Then, (\ref{46}) follows immediately by induction:
\bea
\sum_{t\in {\cal T}^{[n]}}\frac{CM(t)}{t^!}=\sum_{t^\prime\in{\cal T}^{[n-1]}}
\frac{CM(t^\prime)}{(t^\prime)^!}\frac{n-1}{2} & & \\
\Rightarrow
\sum_{t\in {\cal T}^{[n]}}\frac{CM(t)}{t^!}=\frac{(n-2)!}{2^{n-2}}\frac{n-1}{2}=\frac{(n-1)!}{
2^{n-1}}. & &
\eea
To derive (\ref{br}) we write
\be
\sum_{t^{\prime\prime}\in {\cal N}(t^\prime)}
\frac{(t^\prime)^!}{(t^{\prime\prime})^!}
=
\frac{n-1}{n}
\sum_{t^{\prime\prime}\in {\cal N}(t^\prime)}
\frac{\prod (br(t^\prime))^!}{\prod (br(t^{\prime\prime}))^!},
\ee
using the definition of the tree factorial via a product of the
factorials
of the branches.


Natural growth either happens at one of the branches of $t^\prime$,
or at the root of $t^\prime$. In the latter case, the above sum gives a contribution
$\frac{n-1}{n}$. In the former case, assume natural growth happens at the branch
$t^\prime_i$ of $t^\prime$.
Then,
\be
\frac{\prod (br(t^\prime))^!}{\prod (br(t^{\prime\prime}))^!}=\frac{(t_i^\prime)^!}{(
t_i^{\prime\prime})^!},
\ee
where $t_i^{\prime\prime}\in {\cal N}(t_i^\prime)$.
Hence, we can use induction in the above sum,
as branches of a tree have lesser vertices than the tree itself. Thus,
applying (\ref{br}) for branches:
\bea
 & & \frac{n-1}{n}
\sum_{t^{\prime\prime}\in {\cal N}(t^\prime)}
\frac{\prod (br(t^\prime))^!}{\prod (br(t^{\prime\prime}))^!}\\
 & = & \frac{n-1}{n}\left(1+\sum_{i=1}^{f(t^\prime)}\frac{\#(t_i^\prime)}{2}\right)
=\frac{n-1}{n}
\left(1+\frac{\#(t^\prime)-1}{2}\right)=\frac{n-1}{2}.\;\Box
\eea
We used that the sum over all branches of $t^\prime$ delivers $n-2$
vertices, as $t^\prime$ has $n-1$ vertices. Thus, Eq.(46) of
\cite{DiDa} is proven.
\section{Operator Product Expansion}
In a way, the most general problem one faces in QFT
can be stated as the problem of finding the limit
of matrix elements
\bea
 & & \lim_{y\to x}<0\mid O_a(x)O_b(y)P(x_1,\ldots)\mid 0>\nonumber\\
  & &
=\sum_k
f_k(x-y)<0\mid O_k((x+y)/2)P(x_1,\ldots)\mid 0>,\label{vev}
\eea
where $O_a,O_b,O_k$ are operators and $P(x_1,\ldots)$ is a
polynomial
in field operators at points $x_1,\ldots$,
and  we sum over all operators $O_k$ with appropriate
quantum numbers and Wilson
coefficients  $f_k$.
This is the problem of the operator product expansion.
The Wilson coefficients will behave as
\be
f_k(x-y)\sim (x-y)^{-(d_a+d_b-d_k)}(\mbox{\small polynomial in $\log(x-y)$}),
\ee
where $d_a,d_b,d_k$ are the dimensions of the operators
$O_a,O_b,O_k$.


Viewed in momentum space, it becomes an  renormalization
problem: in the desired limit, we will get a series of contributions
which will be plagued by UV divergences when we integrate
momenta. Hence, we can proceed as before and sort the
resulting contributions by their tree-structure, followed
by a renormalization based on the antipode of the resulting trees.
The convolution product extends to the coefficient functions
$f_k$ and one thus finds that the convolution product
of iterated integral is a simple representation of the convolution
which describes the change of scales in OPEs. This motivated the
title of this paper.


The relation between a convolution of the form  (\ref{lll})
and the OPE is most clearly understood if one considers the
OPE as a problem of the change of renormalization conditions,
hence a problem of the change of scales, in our terminology.
It is the standard practice in OPE's to find the desired expansion
by doing the necessary subtractions for the case $x=y$,
which means to find a larger set of appropriate forests
\cite{Collins,H75}.


Let us essentially restrict to $\phi^4$ theory
and consider the problem
\be
\lim_{x\to y}\phi(x)\phi(y)=f(x-y)\phi^2((x+y)/2)+{\cal O}((x-y)^2).
\ee


One starts with the renormalized function
\be
\int d^Dq e^{-iq(x-y)}\Gamma(q,\{p_k\})\label{four}
\ee
associated to the vev (\ref{vev}).
Here,
$\{p_k\}$ indicates the external momenta associated to the points $x_i$
in (\ref{vev}) after Fourier transform, and $q$ is the momentum
according to the Fourier transform of $x-y$.
$\Gamma(q,\{p_k\})$ is a renormalized amputated $(n+2)$-point
Green function
in momentum space (the grey blob
in Fig.(\ref{f4})), dressed with two extra propagators
to connect it to $x$ and $y$.



The corresponding Feynman integrand can be either read as belonging
to a $(n+2)$-point Green function $G^{[n+2]}$ or to a $n$-point
Green function with inserted operator $\phi^2$, $G^{[n]}_{\phi^2}$.
The elements of ${\cal H}$ which we shall
associate to either case will be different.




In the former case, before the limit $x\to y$
is taken, the renormalized
Green function $\Gamma_{G^{[n+2]}}$ is a well-defined finite ratio
for a fixed renormalization scheme $R$.
Now, we want to change this ratio in accord with a new renormalization
condition which demands that the amputated Greens function
vanishes for $\sqrt(q^2)\to \infty$,
so that the momentum integral over $q$ in (\ref{four}) exists.
Note that it would be log-divergent if the amputated Green's function
has a finite value for $q^2\to\infty$.


This is a typical change
achieved by a convolution product described in (\ref{lll}).
The renormalized Green function associated
with the operator insertion, $\Gamma_{G^{[n]}_{\phi^2}}$,
is obtained by convoluting
$\Gamma_{G^{[n+2]}}$ with the ratio which takes into account the change
of renormalization conditions.


Clearly, as the convolution of two ratios is a ratio, we can describe
the so obtained function as a sum over forests.
The leading term $x=y$ is explicitely singled out
in Fig.(\ref{f4}).


Again, a more detailed description of this approach is
mainly a notational exercise and will
be given elsewhere.
\bookfig{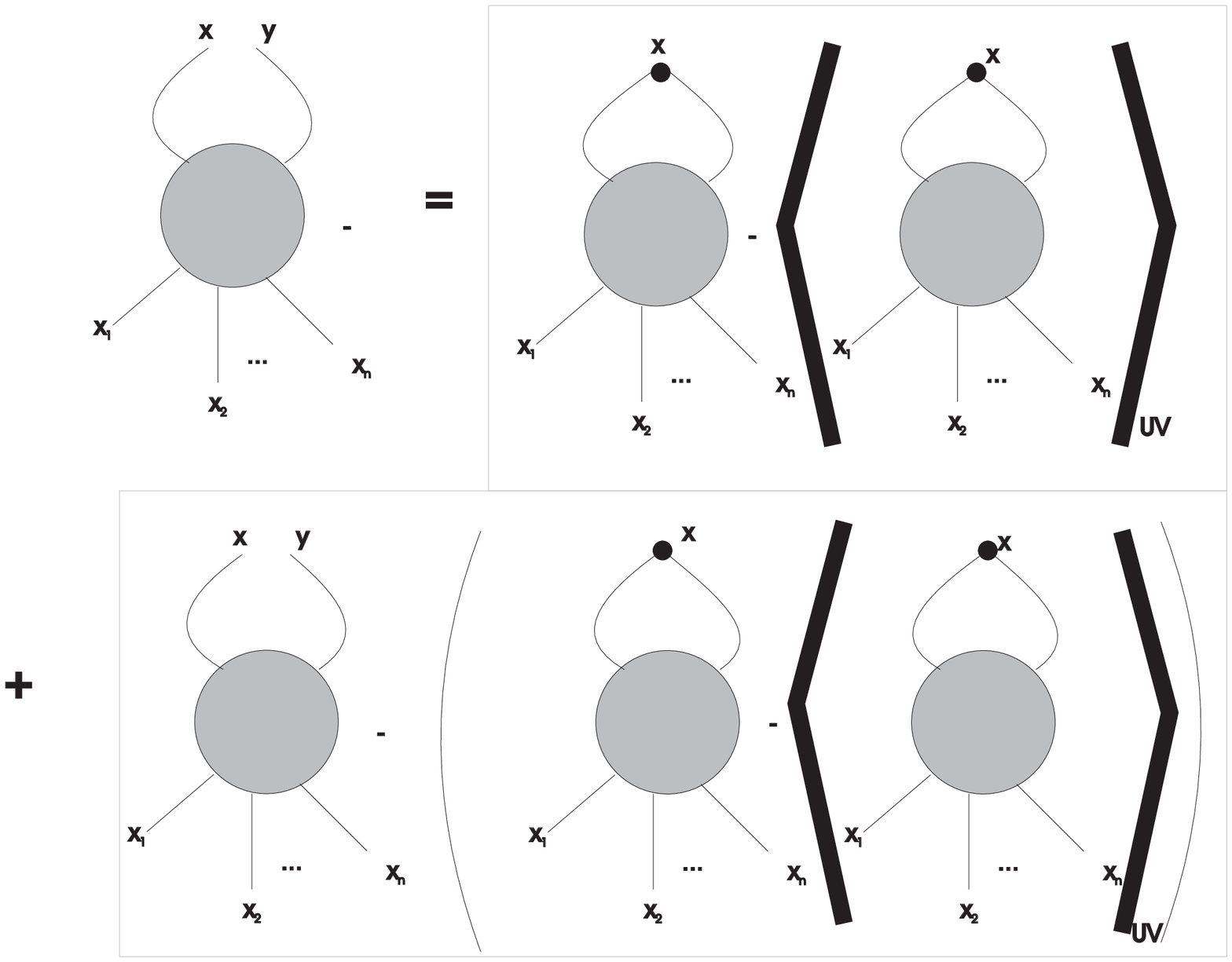}{ope}{f4}{The operator product expansion
amounts to a renormalization of a Green
function such that the limit $x\to y$ is well defined.
It amounts to a change in renormalization conditions
such that the Fourier integral over $q$ can be carried out
for $y\to x$.}{7}
\section{Comments, Conclusions}
In this paper we emphasized the common algebraic structure of
iterated integrals and renormalized Green functions in pQFT. Both
fulfil a convolution law derived from the underlying Hopf algebra
structure. The shuffle product of iterated integrals allows to
restrict the Hopf algebra to ${\cal H}_{Chen}$, while Green
functions represent the full Hopf algebra of decorated rooted
trees. Typical properties of renormalization like RGE's and OPE's
can be derived from this convolution law. The notational exercise
to make this explicit will be presented in future work. The
opposite exercise, expressing QFT Green functions as generalized
iterated integrals, was undertaken in \cite{DKnew}. Here we
presented only the conceptual foundations of renormalization
theory. In the form presented here it inspired already new results
exploring the connection to the Butcher group \cite{Brou} which
also emphasizes the richness of the concepts involved in the
renormalization of a local QFT \cite{CK2}.


We argued that all renormalization schemes can be treated on the same
footing, and gave explicit examples for the case of one-loop
decorations. Two renormalization schemes play a distinguished role:
on-shell schemes (momentum schemes in a massless theory) which serve
as the general mediator for arbitrary changes of the renormalization
point in any choosen scheme and the minimal subtraction scheme which
annihilates scale independent quantities from the beginning.


A hope is that the analytic structure of Green functions
will become coherent in a manner inspired by the study of the
polylogarithm and multiple zeta values
and the structures observed there, ranging
from the K-Z equations and iterated integrals to Hopf
algebras and, eventually, singular knot invariants
and weight systems. The enormous progress in mathematics in recent
years \cite{prog}, so far being mainly related to topological QFT's,
will hopefully enrich our understanding of QFT in four dimensions
eventually. Recent results concerning counterterms of Feynman diagrams \cite{hope1}
as well as analytic structures of Green functions \cite{hope2} justify
some hope, when combined with the results of this paper.


One thing we have not considered here: non-trivial renormalization
schemes establish algebraic structures on $V$ which weaken the structure
of a proper Hopf algebra \cite{hopf}. As this paper is already quite long,
we will describe the resulting weak Hopf algebra structure in more
detail in forthcoming work \cite{KM}.


Let us close this paper with one final observation which shows the
urgent need to work out the connection between renormalization
and NCG \cite{CK} in more detail.
Using the results of section five we write for a rooted tree $t$ with
$n$
vertices
\be
\frac{1}{t^!}=\frac{S_t\,CM(t)}{n!}.
\ee
Plugging this into the model of section four (see \cite{DiDa} to find such
models coming from realistic QFT's), we find
\be
G_z(t;x)=B_t(x)z^{-nx}=\frac{1}{t^!x^{n}}F_t(x)z^{-nx}
\ee
where we factored out the pole parts such that $F(0)=1$.
Summing over all trees we obtain
\be
\sum_n\sum_{t\in {\cal T}^{[n]}}G_z(t;x)=\sum_n \frac{1}{n!}
F_{\delta_n}(x)[\frac{z^{-x}}{x}]^n, \ee where
$F_{\delta_n}(x)=\sum_{t\in{\cal T}^{[n]}}CM(t)S_t F_t(x)$ is the
natural representative of the natural grown $\delta_n$, a nice
result in the light of \cite{CK,CM}, emphasizing that
renormalization  almost (by the deviation of ${\cal R}$ from $id$)
inverts bare Green functions in the group assigned to ${\cal H}$
in the final section of \cite{CK}, in fullagreement with
\cite{Brou} and \cite{CK2}.
\section*{Acknowledgements}
This paper has benefitted enormously from the enthusiasm of and
exchange of ideas with David Broadhurst and Alain Connes. Many
thanks go to Christian Brouder, who had a lot of suggestions on an
earlier version of the ms and made me aware of some literature
relating rooted trees to Runge-Kutta methods. Also, I am grateful
for helpful discussions with Bob Delbourgo, J\"urg Fr\"ohlich,
Reuben Rabi and Ivan Todorov, and it is a pleasure to thank the
latter for hospitality at the Erwin Schr\"odinger Institute in
Vienna, where parts of this paper were conceived. The author is
supported by a Heisenberg Fellowship of DFG.

\end{document}